\documentclass[prx,10pt,twocolumn,superscriptaddress,floatfix,nofootinbib]{revtex4-2}

\usepackage[sc,osf]{mathpazo}
\usepackage[scaled=0.86]{berasans}  
\usepackage[colorlinks=true, allcolors=blue, urlcolor=blue]{hyperref}  
\usepackage{graphicx} 
\usepackage{amsmath,mathtools,amssymb,amsthm,bm,amsfonts,mathrsfs,bbm} 

\usepackage{xspace}  
\usepackage{pgfplots}
\usepackage{xcolor}
\usepackage{bigstrut}
\usepackage{mathrsfs}
\usepackage{dsfont}
\usepackage{tikz}
\usepackage{multirow}
\usepackage{physics}
\usepackage{algpseudocode}

\newcounter{algorithm}
\renewcommand{\thealgorithm}{\arabic{algorithm}}

\newcommand{\algcaption}[2]{%
  \refstepcounter{algorithm}%
  \par\noindent\rule{\linewidth}{0.6pt}\par\vspace{2pt}%
  \noindent\parbox{\linewidth}{\centering\textbf{Algorithm~\thealgorithm. #1}\label{#2}}\par\vspace{2pt}%
  \noindent\rule{\linewidth}{0.4pt}\par\vspace{3pt}%
}

\newcommand{\algfinish}{%
  \par\vspace{2pt}\noindent\rule{\linewidth}{0.6pt}\par
}

\usepackage{caption}
\usepackage{subcaption}
\usepackage{ragged2e}
\DeclareCaptionJustification{justified}{\justifying}
\usepackage{lipsum}

\usepackage{verbatim}

\usepackage{comment}

\theoremstyle{definition}

\theoremstyle{definition}

\theoremstyle{definition}

\theoremstyle{definition}

\theoremstyle{definition}

\theoremstyle{definition}

\newcommand{\bea}{\begin{eqnarray}}
\newcommand{\eea}{\end{eqnarray}}

\newcommand{\I}{\mathds{1}}

\newcommand{\lb}{\nonumber \\}

\newcommand{\id}{\mathsf{id}}
\newcommand{\dg}{^\dagger}

\newcommand{\E}{\mathds{E}}

\newcommand{\G}{\mathds{G}}

\renewcommand{\ket}[1]{|#1\rangle}
\renewcommand{\bra}[1]{\langle#1|}

\usepackage{quantikz}
\newcommand{\nl}{\wireoverride{n}}
\newcommand{\NL}{\setwiretype{n}}
\newcommand{\QL}{\setwiretype{q}}

\begin{document}

\title{Carrier-Assisted Entanglement Purification}

\author{Jaemin Kim}
\email{jaemink@es.aau.dk}
\affiliation{Department of Electronic Systems, Aalborg University, Aalborg 9220, Denmark}

\author{Karthik Mohan}
\email{karthikmohan@kaist.ac.kr}
\affiliation{School of Electrical Engineering, Korea Advanced Institute of Science and Technology (KAIST), Daejeon 34141, Republic of Korea}

\author{Sung Won Yun}
\email{tgb5665@kaist.ac.kr}
\affiliation{Information \& Electronics Research Institute, Korea Advanced Institute of Science and Technology (KAIST), Daejeon 34141, Republic of Korea}

\author{Joonwoo Bae}
\email{joonwoo.bae@kaist.ac.kr}
\affiliation{School of Electrical Engineering, Korea Advanced Institute of Science and Technology (KAIST), Daejeon 34141, Republic of Korea}

\begin{abstract}
Entanglement distillation, a fundamental building block of quantum networks, enables the purification of noisy entangled states shared among distant nodes by local operations and classical communication. Its practical realization presents several technical challenges, including the storage of quantum states in quantum memory and the execution of coherent quantum operations on multiple copies of states within the quantum memory.
In this work, we present an entanglement purification protocol via quantum communication, namely a carrier-assisted entanglement purification protocol, which utilizes two elements only: i) quantum memory for a single-copy entangled state shared by parties and ii) single qubits travelling between parties. We show that the protocol, when single-qubit transmission is noiseless, can purify a noisy entangled state shared by parties. When single-qubit transmission is noisy, the purification relies on types of noisy qubit channels; we characterize Pauli channels such that the protocol works for the purification. We address this limitation by using multiple carrier qubits, and show that for any non-entanglement-breaking Pauli channel, the protocol's fixed-point fidelity approaches unity as the number of carriers increases.
Our results significantly reduce the experimental overhead required for distilling entanglement: the practical advantage is demonstrated through parameters directly related to the capability of entanglement purification, such as noise in quantum memory, local measurements, channel use, and entanglement fidelity. We envisage that the protocol would make long-distance pure entanglement closer to a practical realization.
\end{abstract}


\maketitle

\section{Introduction}

Information processing based on the laws of quantum mechanics, such as quantum computing and quantum communication, enables efficient computation, enhanced security, and improved communication capabilities beyond the limitations of current information technologies. An entangled bit (ebit), the unit of entanglement, is a key resource for quantum information processing,
\bea
|\phi^+\rangle =\frac{1}{\sqrt{2}}(|00\rangle + |11\rangle). \label{eq:ebit}
\eea
Extensive efforts have been devoted to devising methods for distributing pure entangled states in practical scenarios. An entanglement purification protocol (EPP) \cite{bennett1996purification, Deutsch1996} begins with multiple pairs of noisy entangled states and sacrifices a significant fraction of them to distill ebits, using local operations and classical communication (LOCC).

{Several EPPs have been proposed. The two seminal protocols, BBPSSW~\cite{bennett1996purification} and DEJMPS~\cite{Deutsch1996}, have been generalized \cite{Fujii2009, Dur:2007aa} and also extended to take measurement errors into account ~\cite{kim2025purification}.}
A more practical scheme, called entanglement pumping (EP) \cite{Dur:2007aa}, which can be interpreted as two-way entanglement activation, i.e., probabilistic entanglement enhancement \cite{kim2025detectingentanglementstatepreparation}, has also been proposed. Note that all these are two-way distillation protocols. Enhancement of entanglement by two-way EPPs has been experimentally demonstrated \cite{pan2003experimental, Reichle2006_Nature_Ions, Chen2017_NatPhoton_Nested, yan2022entanglement, siddhu2025basic, Zhou:20, kbw2-fdqn, PhysRevLett.126.010503}. There are also one-way EPPs such as the hashing protocol \cite{Bennett1996Mixed}. Comparing one- and two-way protocols, one can find that one-way EPPs are more efficient for sufficiently highly entangled states, and two-way EPPs work for a wider range of entangled states as they tolerate highly mixed entangled states. EPPs for multipartite entanglement, such as graph states and cluster states, which are particularly useful for quantum computing, have also been proposed~\cite{Dur2003MultiparticleGraphStates, Aschauer2005TwoColorableGraphStates}.  

The challenge in implementing EPPs for practical applications lies in the technology requirements for their realization, specifically quantum memory and local quantum operations. Parties can store quantum states by using quantum memory in order to apply local quantum operations over multiple pairs of a shared state. To distill an ebit at the end, an EPP needs a substantially large demand on the storage capacity~\cite{Collins2007_MultiplexedRepeaters,Rozpedek2018Optimizing}. Moreover, noise in local operations consisting of quantum gates, measurements, and quantum memories further degrades the entanglement fidelity~\cite{briegel1998quantum}. One can notice that the state-of-the-art experimental demonstrations have shown a few rounds of an EPP~\cite{pan2003experimental,Reichle2006_Nature_Ions,Chen2017_NatPhoton_Nested,yan2022entanglement,siddhu2025basic}, and an EPP with multi-rounds for practical purposes remains unachievable yet.

\begin{figure*}[t]
\centering
\begin{quantikz}[every arrow/.append style={no head}]
\nl & \nl \lstick{\footnotesize $A_0$} & \gate{U_A} & \ctrl{1} & & & & & & & & \rstick[5]{$\rho_{\mathrm{out}}$}\\
\nl & \nl & \nl \midstick{Carrier: $\ket{0}$} & \targ{} & \gate{V_A} & \permute{3}  \\
\lstick{$\rho$} \NL &&&&& \gate{\mathcal{N}} &&& \\
\NL & & & & & & \gate{V_B} \wire[l][1][shorten >=2.5mm]{q} & \targ{} \QL& \meter{0} \\
\nl & \nl \lstick{\footnotesize $B_0$} & \gate{U_B} & & & & & \ctrl{-1} & & & &
\arrow[from=3-1, to=1-2, line width=0.9pt]
\arrow[from=3-1, to=5-2, line width=0.9pt]
\end{quantikz}
\caption{ 
The CAEPP with a single-qubit carrier is shown. When two parties share a state $\rho$, Alice prepares an ancilla qubit in a state $|0\rangle$, applies a CNOT gate as an encoding, and sends a carrier qubit to Bob, who applies a CNOT gate. Once a measurement outcome on a carrier qubit is $0$, two parties accept a resulting state, denoted by $\rho_{\mathrm{out}}$.
{When a channel for a carrier qubit is noisy, denoted by $\mathcal{N}$, two parties may implement encoding and decoding strategies for carrier qubits. Carrier coding with local unitaries, denoted by $V_A$ and $V_B$, is detailed in the sections \ref{sec:manipulate} and \ref{sec:channel_twirling}.} } 
\label{fig:carrier}
\end{figure*}

On the one hand, mature quantum technologies will be useful as they can realize long-term quantum memory and fault-tolerant coherent quantum state manipulation. On the other hand, it is also important to devise a practical EPP with minimal technology requirements. For instance, quantum memories remain expensive resources, and noisy quantum operations will increase errors. Therefore, an EPP with minimal experimental resources, such as quantum memory, quantum gates, and measurements, leads the ultimate ebit-connected network closer to reality.

We here propose a novel and practical protocol for purifying noisy entanglement through noisy quantum communication without consuming additional entangled pairs, called the carrier-assisted entanglement purification protocol (CAEPP). {Technically, it may be rephrased as a memory-lean implementation of entanglement purification using transmitted carrier qubits instead of stored noisy entangled pairs.} The protocol is practical as experimental requirements are minimal {in terms of parameters directly related to the capability of entanglement purification}: two quantum memories for a single copy of a two-qubit entangled state and a channel for single-qubit transmission. Compared to two-way EPPs (TWEPPs), the protocol is more robust against measurement errors as it performs fewer measurements. It is novel in that an ebit can be ultimately distilled by exploiting multiple-qubit transmission, which we formulate as multi-carrier-assisted EPP (mCAEPP); the result also improves EP so that EP not only realizes an activation of entanglement but also reaches the distillation of an ebit. Thus, the protocol efficiently realizes an increase of entanglement with the aforementioned minimal experimental resources. 
The CAEPP can be extended to multipartite systems, such as the purification of {GHZ} states.


The article is structured as follows. In Section \ref{sec:preliminaries}, terminologies and notations are briefly summarized. In Section \ref{sec:CAEPP}, we introduce the CAEPP in the simplest form to deliver the central idea of the protocol and provide examples when a single-qubit carrier is noiseless. In Section \ref{sec:performance}, we consider a realistic scenario where a single-qubit carrier suffers from noisy transmission. We show that a noisy carrier can increase entanglement of a shared state up to a threshold, called the maximum convergent fidelity. In Section \ref{sec:multiple_carrier}, we resolve the limitation by multiple carriers: namely, we show that the CAEPP through noisy channels can distill ebits by exploiting multiple-qubit carriers. In Section \ref{sec:comparison}, we present comparisons of the CAEPP and TWEPPs through experimental resources and the effects of noise in implementation. In Section \ref{sec:GHZ}, 
we demonstrate that the CAEPP can be generalized to purification for the GHZ state.
In Section \ref{sec:conclusion}, we summarize the results and discuss future directions.

\section{Preliminaries} \label{sec:preliminaries}

Let us begin with a summary of notations and terminologies to be used throughout. Let $\phi^{\pm}$ and $\psi^{\pm}$ denote Bell states as follows,
\begin{align*}
\ket{\phi^\pm}=\tfrac{1}{\sqrt{2}}(\ket{00}\pm\ket{11}),\qquad
\ket{\psi^\pm}=\tfrac{1}{\sqrt{2}}(\ket{01}\pm\ket{10}).
\end{align*}
We write $\phi^\pm := \ket{\phi^\pm}\!\bra{\phi^\pm}$ and $\psi^\pm := \ket{\psi^\pm}\!\bra{\psi^\pm}$.

{ A Pauli channel} connecting Alice and Bob, denoted by $\mathcal{N}$, can be characterized by probabilities $(p_{00},p_{01},p_{10},p_{11})$ satisfying $\sum_{ab} p_{ab}=1$ as follows,
\begin{align}
\mathcal{N}(\cdot)=p_{00}(\cdot)+p_{01}\,Z(\cdot)Z+p_{10}\,X(\cdot)X+p_{11}\,Y(\cdot)Y, 
\label{pauli_chan}
\end{align}
where $X,Y,Z$ are the Pauli operators. { Note also that channel twirling with Pauli matrices can transform a single-qubit channel into a Pauli channel \cite{Dur2005}. Note also that a channel $\mathcal{N}$ is called entanglement-breaking (EB) if resulting states $[\id\otimes \mathcal{N}](\sigma)$ are separable for all bipartite states $\sigma$ \cite{doi:10.1142/S0129055X03001709}. In fact, it holds that $\mathcal{N}$ is EB if $[\id\otimes \mathcal{N}] (| \phi^+\rangle \langle \phi^+|)$ is separable \cite{doi:10.1142/S0129055X03001709}; applying this condition to Eq. (\ref{pauli_chan}), one can find that $\mathcal{N}$ is EB if and only if $\max\{p_{ij}\} \le 1/2$.
} 

For convenience, throughout, we assume that probabilities are in a decreasing order, i.e.
\bea
p_{00} \geq p_{01} \geq p_{10} \geq p_{11}.  \label{eq:order}
\eea
Note that local unitaries can realize permutations of probabilities $p_{ij}$ \cite{Acin2006}, i.e.,
\bea
(p_{00},p_{01},p_{10},p_{11}) \leftrightarrow (p_{\pi(00)},p_{\pi(01)},p_{\pi(10)},p_{\pi(11)}) \nonumber
\eea
for some permutation $\pi$.

{ Suppose that Alice prepares a state $|\phi^+\rangle$ and sends the second qubit to Bob through a Pauli channel in Eq. (\ref{pauli_chan}), then two parties share a mixture of four Bell states with the condition in Eq. \eqref{eq:order}, denoted by $\rho$. The entanglement fidelity of a two-qubit state is defined as follows,
\begin{align}
F(\rho) := \bra{\phi^+}\rho\ket{\phi^+} = p_{00},
\label{eq:entf}
\end{align}
can be used to quantify entanglement. For instance, an ebit can be identified by $F = 1$. Note that a two-qubit state is entangled, if $F>1/2$ \cite{Bennett1996Mixed, PhysRevLett.77.1413, Horodecki:1996aa}. }  



\section{Carrier-Assisted Entanglement Purification Protocol} \label{sec:CAEPP}

In this section, we present the carrier-assisted entanglement purification protocol (CAEPP).

\subsection{The protocol}
\label{subsec:protocol}

To begin the protocol, two parties, Alice and Bob, establish a single copy of a two-qubit state between them,
\begin{align}
\rho \;=\; q_{00}\phi^+ + q_{01}\phi^- + q_{10}\psi^+ + q_{11}\psi^-,
\label{qBDS}
\end{align}
where the fidelity is given by $F=q_{00}$, see also \textbf{Initialization} in {\bf Algorithm} \ref{CAEPP}. Note also that a shared state satisfies the order among probabilities $\{q_{ij}\}$ in \eqref{eq:order} such that $q_{11}$ is the minimal one.

The steps of a single round in the CAEPP are in order.  
\begin{enumerate}
\item \textbf{Pre-processing:} Two parties apply local operations $R_X(+\tfrac{\pi}{2})\otimes R_X(-\tfrac{\pi}{2})$ to a shared state, so that a state $\phi^-$ in a shared Bell-diagonal state is the least probable, i.e., $q_{01}$ is the minimal one. 
\item \textbf{Encoding:} Alice prepares a carrier qubit in a state $\ket{0}$. She applies a CNOT gate by taking the carrier as a target qubit, and sends the carrier through a channel $\mathcal{N}$.
\item \textbf{Decoding:} Once Bob receives a qubit from Alice, he applies a CNOT gate by taking the carrier as a target qubit. A measurement is performed on the carrier in the $Z$ basis.
\item \textbf{Decision:} Bob tells Alice a measurement outcome. If an outcome is $0$, Bob declares \textbf{Success} and keeps the shared pair; otherwise, he announces \textbf{Failure} to discard a shared state.
\end{enumerate}

In the CAEPP, {\bf Pre-processing} realizes a Bell-diagonal state with $q_{01}$ minimal. {\bf Encoding} corresponds to an entangling gate at Alice. Then, $X$ and $Y$ errors that may happen during transmission can be detected in {\bf Decoding} and discarded in {\bf Decision} by Bob.
The other type of errors $Z$ will be arranged and considered in the next round {\bf Pre-processing}. The protocol can be generalized with multiple carriers $m\geq 1$, where stabilizer purification protocols apply, see Algorithm~\ref{CAEPP}.

{ Let us also rephrase that the CAEPP exploits two channels, one for sharing an entangled pair that is stored in quantum memories of two parties, and the other for carrier-qubit transmission. In general, and also in practice, two channels may have different levels of noise. Moreover, noise in an entangled pair may not have its origin from a channel but also quantum memory; in this case, the effect of noise on a shared state is replaced by memories from a channel. Hence, the CAEPP allows two parties to maximally exploit available experimental resources, two memories and single-qubit transmission, to increase or distill entanglement. The scenario may be compared with EP as follows.} 

{ {\it Comparison to EP:} It is worth mentioning an EP protocol \cite{Dur:2007aa}, see also recent experimental realization \cite{doi:10.1126/sciadv.ado9822}. Let us recall that an EP protocol exploits copies of entangled states, called elementary pairs, having a fidelity $F_0$. Contrasting to EPPs in \cite{bennett1996purification, Deutsch1996}, it repeatedly introduces an elementary pair to increase the fidelity of a fixed one, which may then improve from $F_0$ to a threshold $F_*$, where $F_*<1$ unless an elementary pair has $F_0=1$. An EP protocol shares some similarity with the CAEPP in that only two quantum memories are needed, which are for a fixed pair. Mathematically, i.e., when local operations are noiseless, an EP protocol is equivalent to the CAEPP whenever an elementary pair defines a noisy channel for a single-qubit carrier through the well-known state-channel duality, the Choi--Jamiołkowski (CJ) isomorphism
\cite{Choi1975, Jamiolkowski1972}.
In this case, as to be described, $F_*$ maximally achievable in an EP protocol is equal to $F_{\star}$ in Eq. (\ref{eq:fstar}). 

In practical realizations, an EP protocol applies two measurements, one of which is replaced by the preparation of a carrier qubit in the CAEPP. Hence, compared to EP, the CAEPP is less vulnerable to measurement imperfections; in practice, single-photon measurements often exhibit higher error rates than state preparation. Moreover, as we extend to $m$-qubit carriers, the CAEPP will ultimately reach the purification $F_{\star}\rightarrow 1$, which may introduce collective EP: that is, pumping with collective pairs can resolve the limitation of an EP protocol to reach the complete purification, i.e. $F_*\rightarrow 1$. In contrast with the CAEPP, EP with multiple copies may not be feasible without a further requirement on quantum memory. Thus, we summarize that the CAEPP is implementation-friendly and, on top of that, as we shall describe, presents a prescription to EP to reach the maximal fidelity $F_{\star} \rightarrow 1$ through manipulations of collective copies. }



\subsection{Two-round purification via noiseless transmission}

Suppose that a channel $\mathcal{N}$ for a single-qubit carrier is noiseless, that is, $\mathcal{N}=\id$. 
{ For simplicity, assume Alice and Bob omit the \textbf{Pre-processing} at the first round.}
Following the CAEPP, we find that once Alice and Bob accept a shared state after the protocol, the resulting state is given by, 
\begin{align}
\rho_{\mathrm{out}} \;=\; q_{00}'\phi^+ + q_{01}'\phi^- + q_{10}'\psi^+ + q_{11}'\psi^-, \label{eq:rho_out}
\end{align}
where
\bea
q_{00}' &= & \frac{q_{00}}{q_{00} + q_{01}},~~~
q_{01}'= \frac{ q_{01}   }{q_{00} + q_{01}} \nonumber 
\eea
and 
\bea
q_{10}' =q_{11}'=0 \nonumber
\eea
One can observe that the protocol accepts Bell states $\phi^{\pm}$ with an even parity and rejects the others $\psi^{\pm}$ having an odd parity. 

Afterwards, two parties run the protocol for the state $\rho_{\mathrm{out}}$ in \eqref{eq:rho_out}. After the pre-processing step, two parties share a mixture of two Bell states $\phi^+$ and $\psi^-$. When Bob announces \textbf{Success} after the protocol, which successfully rejects a Bell state $\psi^-$ having an odd parity, a resulting state is $\phi^+$ and thus $q_{00}''=1$, which has a fidelity $F=1$. 

We have shown that two successful rounds of the CAEPP through noiseless single-qubit transmission purify noisy entanglement and enable two parties to share an ebit.




\begin{figure*}[t]
\begin{minipage}{0.45\textwidth}
\centering
\includegraphics[width=\linewidth]{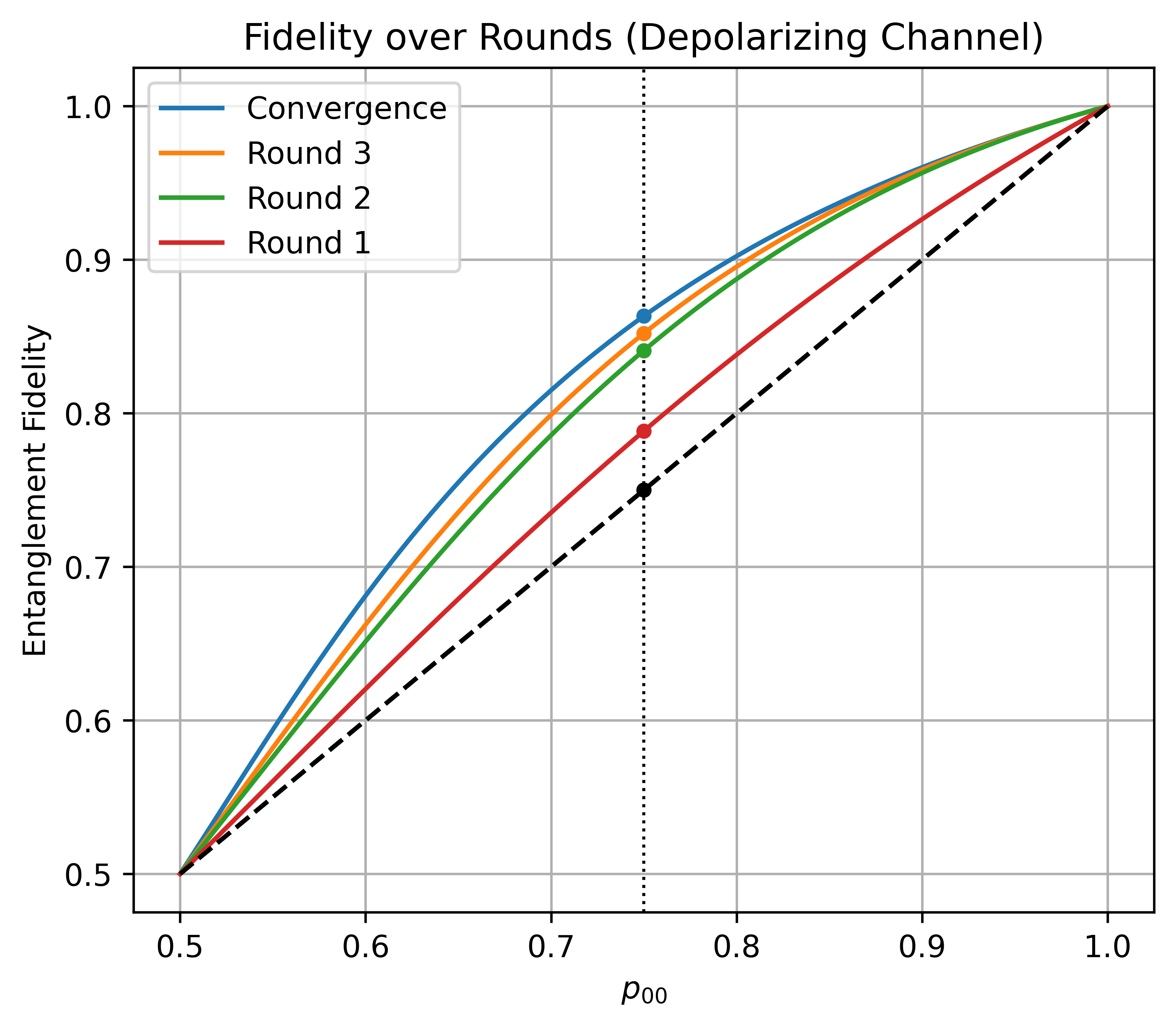}
\end{minipage}\hfill
\begin{minipage}{0.45\textwidth}
\centering
\includegraphics[width=\linewidth]{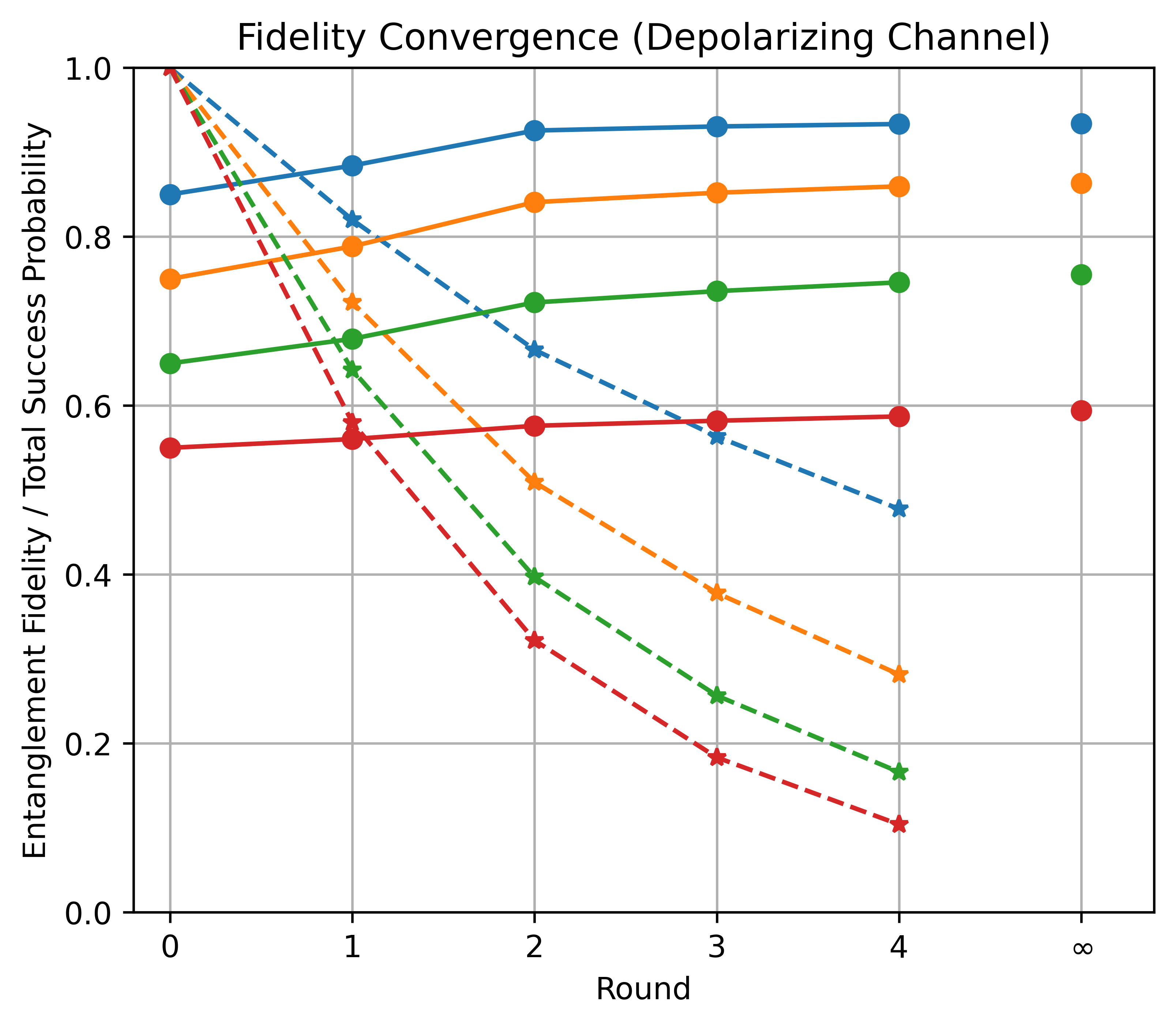}
\end{minipage}
\vspace{-0.6em}
\caption{{ The CAEPP has two channels, one for transmitting a carrier against a noisy channel, and the other for sharing an entangled pair. For fair comparisons with TWEPPs, let us assume that the two channels are identical and that both are depolarizing channels. (a) The fidelity of a shared entangled state increases by a single-qubit carrier. For instance, a pair with $F=p_{00}=0.75$ (black dot), a single round of the CAEPP increases it to $0.788$ (red dot), and then the next one to $0.841$ (green dot). The fidelity converges to $F_{\star}\approx 0.863$. (b) The entanglement fidelity (solid line) increases quickly so that it is sufficiently close to the maximal convergent fidelity after four rounds. The first two rounds are more efficient than the next two rounds. The total success probability (dotted line) of accepting an entangled pair decreases. For instance, when an initial fidelity is $F=0.75$, the total success probability after four rounds is about $0.3$ (orange line).} }
\label{fig:depol}
\end{figure*}

{ \subsection{Carrier encoding against noisy transmission I: the deterministic strategy} 
\label{sec:manipulate}}

We revisit the step \textbf{Pre-processing} that applies local unitaries and discuss effective encoding on a single-qubit carrier against a noisy channel, see also Fig. \ref{fig:carrier}. We recall that the step applies local unitaries $ R_X\left(+\tfrac{\pi}{2}\right) \otimes R_X\left(-\tfrac{\pi}{2}\right)$ to a Bell-diagonal state in \eqref{qBDS} that satisfies the order in \eqref{eq:order}. An application of the local unitaries realizes a permutation among error probabilities, 
\bea
(p_{00},p_{01},p_{10},p_{11}) \mapsto (p_{00},p_{11},p_{10},p_{01}) \label{eq:perm}
\eea
so that $\phi^-$ is the least probable while the others remain unaltered. Then, each round of the CAEPP effectively suppresses errors of odd parities $\psi^{\pm}$. 

From the duality between states and channels, called the CJ isomorphism, the permutation in \eqref{eq:perm} can be realized on a channel. In general, via the CJ isomorphism, a transformation of bipartite states by local unitaries $U\otimes V$ is equivalent to manipulating a channel $\mathcal{N}$ such that 
\bea
V\mathcal{N}(U^{T} (~\cdot ~) U^{*}) V^{\dagger}. \label{eq:cht}
\eea
The expression above shows that a sender applies a unitary $U^{T}$ as an encoding on a state before a channel, and a receiver performs a unitary $V$ as a decoding after receiving a state.

Then, the permutation in \eqref{eq:perm} can be realized for a Pauli channel $\mathcal{N}$ as 
\bea
R_X\left(+\tfrac{\pi}{2}\right) ~ \mathcal{N} (R_X\left(-\tfrac{\pi}{2}\right) (\cdot) R_X\left( \tfrac{\pi}{2}\right) ) ~R_X\left(- \tfrac{\pi}{2}\right). \label{eq:pi/2}
\eea
Hence, we can identify an encoding before a channel use and a decoding after the channel as follows,
\begin{itemize}
\item {carrier} encoding: Alice applies a rotation gate $R_X\left(-\tfrac{\pi}{2}\right) $ before sending, and
\item {carrier} decoding: Bob implements a rotation gate $R_X\left(+\tfrac{\pi}{2}\right) $ before a measurement. 
\end{itemize}
The encoding and decoding strategy of two parties shuffles Pauli errors $Z$ and $Y$ in a Pauli channel, equivalently to the permutation in \eqref{eq:perm}.

For instance, let us consider a Pauli channel with four probabilities,
\bea
\bigl(p_{00}, \tfrac{2}{5}(1-p_{00}), \tfrac{2}{5}(1-p_{00}), \tfrac{1}{5}(1-p_{00})\bigr). 
\label{biased_chan}
\eea
Applying the encoding and decoding strategy, two parties are effectively connected by some other Pauli channel 
\bea
\bigl(p_{00}, \tfrac{1}{5}(1-p_{00}), \tfrac{2}{5}(1-p_{00}), \tfrac{2}{5}(1-p_{00})\bigr),
\label{new_biased_chan}
\eea
where the fraction of a $Z$-error is minimal. 

Hence, the CAEPP can efficiently suppress the largest fractions of errors causing odd parities $X$ and $Y$. The $Z$ and $Y$ errors are permuted by the encoding and decoding strategy and eliminated by the CAEPP. After all, the CAEPP can suppress the fractions of the error rates of all types, leading to distilling $\phi^+$.

{ 
\subsection{Carrier encoding against noisy transmission II: the random-unitary strategy} 
\label{sec:channel_twirling}

The encoding and decoding for carrier qubits in Eq. (\ref{eq:cht}) can be generalized with unitary $2$-design, a collection of unitaries $\mathbbm{U}= \{U_i \}_{i=1}$ that fulfills a state transformation as follows, for a two-qubit state $\rho$, 
\begin{align}
\frac{1}{ |\mathbbm{U}| }\sum_i U_i \otimes U_{i}^{*} \rho (U_i \otimes U_{i}^{*} )^{\dagger} = \alpha|\phi^{+}\rangle\langle \phi^+| + \frac{1-\alpha}{4} \I \nonumber
\end{align}
where $\alpha$ is related to the entanglement fidelity of an initial state $\rho$, $F=\langle\phi^+|\rho|\phi^+\rangle$, which is invariant under twirling \cite{PhysRevA.60.1888}. Thus, it holds that $\alpha = (4F-1)/3$. Note that twirling above is precisely what has been exploited in the BBPSSW protocol \cite{bennett1996purification, PhysRevA.80.012304}. Then, following twirling with unitary $2$-design, Alice sends a state after encoding that applies a unitary $U_{i}^T$ and Bob implements $U_{i}^{*}$ once he receives the state, for a randomly chosen unitary $U_i\in \mathbbm{U}$. It holds that for a channel $\mathcal{N}$ connecting Alice and Bob, 
\begin{align}
\frac{1}{ |\mathbbm{U}| } \sum_{i} U_{i}^{*} \mathcal{N} ( U_i^T (\cdot) U_{i}^{*})  U_{i}^T 
= \frac{4F-1}{3} ~\mathrm{id}(\cdot) + \frac{4-4F}{3} D(\cdot). \label{eq:red}  
\end{align}
where $\mathrm{id}$ is an identity map and $D$ denotes a complete depolarizing channel, $D(\rho) =\I/2$ for all qubit states $\rho$. Hence, random encoding in Eq.~(\ref{eq:red}) establishes a depolarizing channel for an arbitrary channel between two parties. 

Two parties apply the technique of random encoding to an arbitrary and unknown Pauli channel \cite{PhysRevA.99.062302, 10993387}, and can realize a channel transformation as follows, 
\[
(p_{00},\ p_{01},\ p_{10},\ p_{11}) \mapsto (p_{00},\ \frac{1-p_{00}}{3},\ \frac{1-p_{00}}{3},\ \frac{1-p_{00}}{3})
\]
so that they establish a depolarizing channel. Note that the no-error probability \(p_{00}\) remains unaltered \cite{Bennett1996Mixed,Dur2005}. Note that random encoding has been used to preserve optimal measurements in quantum communication over noisy channels \cite{PhysRevA.99.062302, 10993387}. 

We remark that the technique of random encoding for carrier qubits, which establishes a depolarizing channel, can be replaced with the scheme with a fixed rotation in Eq. (\ref{eq:cht}). For instance, the proposed random encoding and decoding strategy implements a transformation for a Pauli channel in Eq. (\ref{biased_chan})
\bea
\bigl(p_{00}, \tfrac{1}{3}(1-p_{00}), \tfrac{1}{3}(1-p_{00}), \tfrac{1}{3}(1-p_{00})\bigr), \nonumber
\eea
whereas the strategy in Eq. (\ref{eq:pi/2}) leads to another Pauli channel in Eq. (\ref{new_biased_chan}). 
}

\begin{figure*}[t]
\begin{minipage}{0.45\textwidth}
\centering
\includegraphics[width=\linewidth]{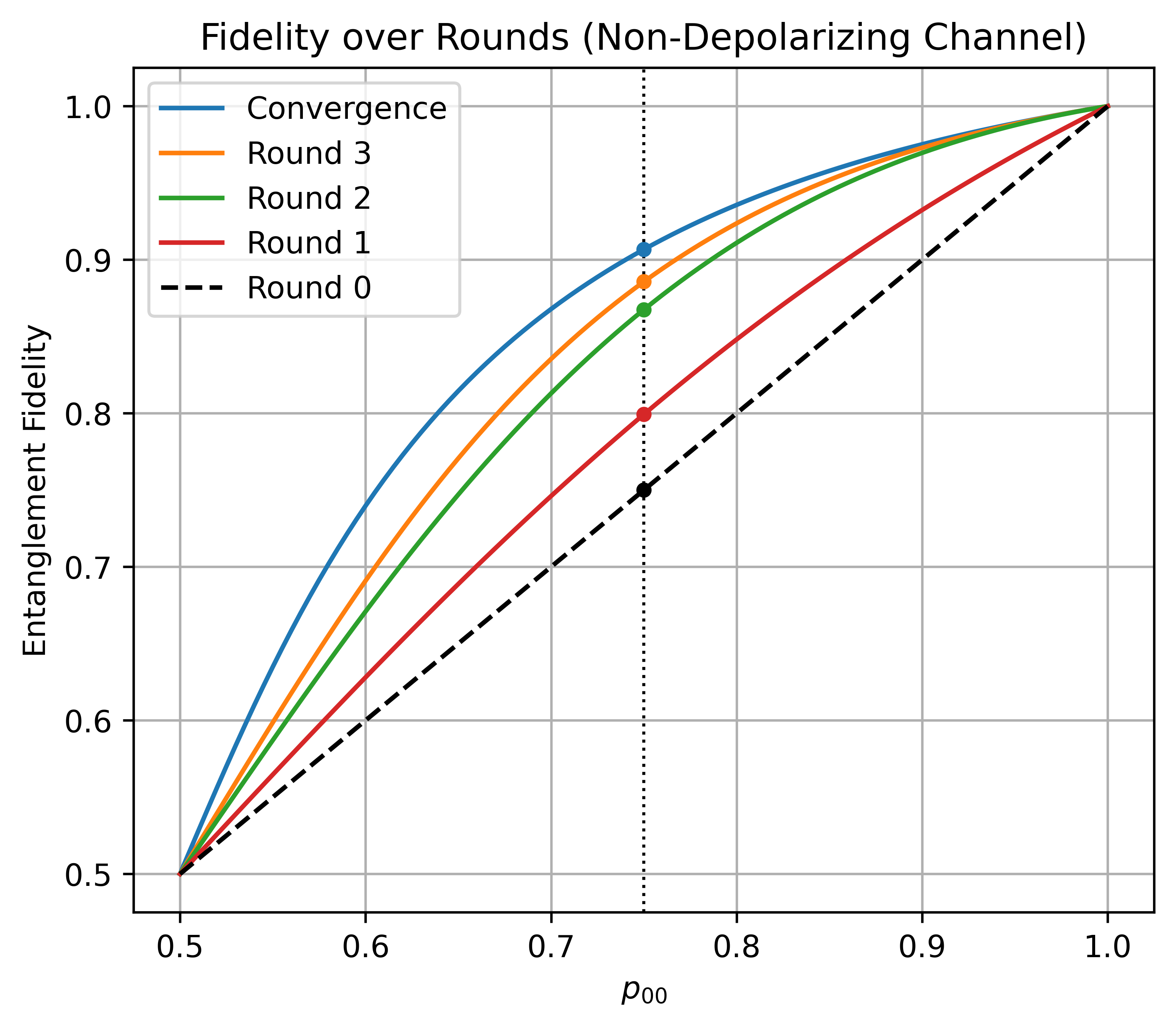}
\end{minipage}\hfill
\begin{minipage}{0.45\textwidth}
\centering
\includegraphics[width=\linewidth]{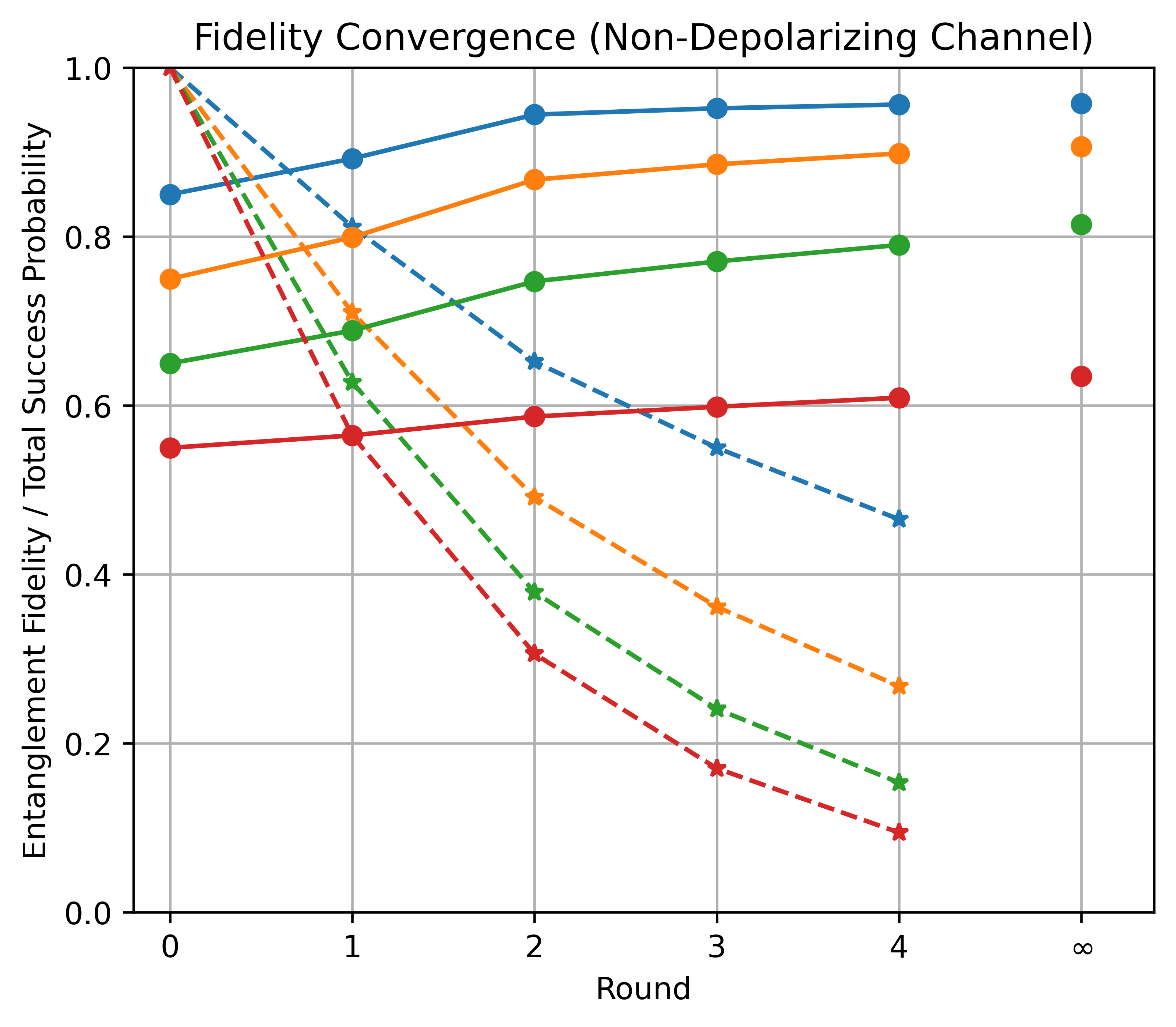}
\end{minipage}
\vspace{-0.6em}
\caption{ { The CAEPP exploits a noisy channel, a general Pauli channel in \eqref{new_biased_chan}, to share an entangled pair and transmit a single-qubit carrier. Compared to the case of depolarizing channels in Fig. \ref{fig:depol}, one can find that the maximum convergent fidelity is higher as the $Z$-error rate is lower. (Left) The entanglement fidelity increases by exploiting single-qubit carriers. (Right) Four rounds of the CAEPP reach an entanglement fidelity (solid line), sufficiently close to the maximal convergent fidelity $F_{\star}$. The total success probability decreases as the CAEPP applies more single-qubit carriers. For instance, for an initial fidelity $F=0.85$, the CAEPP increases it up to $F=0.95$ after four rounds. The total success probability is about $0.5$ (blue line). }
} 
\label{fig:biased_channel}
\end{figure*}

{ 
\subsection{Bell-diagonal coefficient update and fidelity gain condition}
Suppose a shared state at the start of a round, i.e., before \textbf{Pre-processing}, as follows, 
\begin{align}
\rho = q_{00}\phi^+ + q_{01}\phi^- + q_{10}\psi^+ + q_{11}\psi^-, \nonumber
\end{align}
which can be characterized by a vector $\bm{q} = (q_{00}, q_{01}, q_{10}, q_{11})^\top$. A single purification round with a Pauli channel with parameters $(p_{00}, p_{01}, p_{10}, p_{11})$, together with the steps of \textbf{Pre-processing}, \textbf{Encoding}, \textbf{Decoding}, and \textbf{Success} at \textbf{Decision}, leads to a resulting Bell-diagonal state denoted by $\rho'$ characterized by $\bm{q'} = (q'_{00}, q'_{01}, q'_{10}, q'_{11})^\top$ given by
\begin{align}
\bm{q'} &= \frac{1 }{P_\mathrm{succ}}L \bm q, 
\quad 
L = 
\begin{bmatrix}
p_{00} & 0 & 0 & p_{01}\\
p_{01} & 0 & 0 & p_{00}\\
0 & p_{11} & p_{10} & 0\\
0 & p_{10} & p_{11} & 0
\end{bmatrix}, \label{eq:coeff_update}
\end{align}
where $P_\mathrm{succ} = \bm{1}^\top L \bm q$ denotes the success probability and $\bm{1} = (1,1,1,1)^\top$. 
One can find that the entanglement fidelity increases if $q_{00}'>q_{00}$: 
\begin{equation}   
\frac{q_{00} p_{00} + q_{11} p_{01}}{ (q_{00} + q_{11}) (p_{00} + p_{01}) + (q_{01} + q_{10})(p_{10} + p_{11}) } > q_{00}. \label{eq:condition}
\end{equation}
Hence, the general condition relies on two collections of parameters, one from a shared state and the other from a channel for single-qubit carrier. Note that the results of TWEPPs on two isotropic copies can be seen as an instance when a channel and a shared state are equally noisy, i.e., $\bm{q} = \bm{p}$.
}

\section{CAEPP with noisy single-qubit carriers} \label{sec:performance}

We here consider noisy channels for a single-qubit carrier, that is, the CAEPP with noisy single-qubit carriers. We assume that a noisy channel $\mathcal{N}$ for a carrier is identical to the channel defining a shared state. The assumption follows from a realistic constraint; to establish a shared state, Alice creates an ebit and sends a qubit to Bob, where a qubit from Alice has also experienced an identical noisy channel, through which a carrier is also to be transmitted. 


\subsubsection*{The scenario}
As mentioned, both qubits, one for establishing a shared state and the other for a single-qubit carrier, are sent over an identical noisy channel. The scenario can also be rephrased as two copies of channels $\mathcal{N}\otimes \mathcal{N}$ where one is for sharing an entangled state and the other is for a single-qubit carrier. The CAEPP applies, along with the encoding and decoding strategy explained in the subsection \ref{sec:manipulate}. 

In the step \textbf{Initialization}, Alice prepares $\phi^+$ and sends one qubit to Bob through a channel $\mathcal{N}$, so a Bell-diagonal state in ~\eqref{qBDS} is shared. After the protocol, once Bob obtains an outcome $0$ and announces \textbf{Success}, a state shared by two parties has a higher purity, being closer to an ebit. 

\subsubsection*{Remark} 
A shared entangled state may have a higher fidelity by the CAEPP with noisy transmission of a single-qubit carrier. However, it cannot be purified to an ebit in general if a carrier qubit suffers from a noisy channel. We write by
\bea
F_{\star}:\mathrm{~the ~maximum ~convergent ~fidelity}  \label{eq:fstar}
\eea
of a resulting shared state from repetitions of the CAEPP. 


{
Successful rounds of the CAEPP suppress the fraction of states $\psi^\pm$; the pre-processing in Eq.~(\ref{eq:order}) is chosen to minimize the fraction of a state $\phi^-$. Overall, repeating the CAEPP can suppress fractions of all Bell states but a desired one $\phi^+$.

In what follows, we begin with a depolarizing channel to elucidate the CAEPP and then consider a Pauli channel with different coefficients. If a Pauli channel is not of full rank, two parties can re-order the coefficients such that $\phi^-$ is not present, see Eq. (\ref{eq:order}). In this case, it holds that the convergent fidelity reaches $F_{\star}=1$. 
}

\subsection{Pauli channels: $F_{\star}<1$}

A Pauli channel is characterized by four probabilities $\{p_{ij} \}$, and let us consider general cases where all of them are nonzero.


{Let us begin by considering an illuminating example with a depolarizing channel: the channel has equal probabilities for errors $X$, $Y$, and $Z$}, for which we parameterize probabilities characterizing a channel as follows, 
\bea
\bigl(p_{00},\,\tfrac{1}{3}(1-p_{00}),\,\tfrac{1}{3}(1-p_{00}),\,\tfrac{1}{3}(1-p_{00})\bigr). \label{depol_chan}
\eea
Note that through the channel, when Alice creates an ebit and sends one qubit to Bob, two parties share an isotropic state, 
\bea
\rho_{\mathrm{iso}} = p_{00}\,\phi^+ + \frac{1-p_{00}}{3}\,(\I-\phi^+).
\label{iso}
\eea
which is entangled for $p_{00}>1/2$. 

{Then, repeating the CAEPP through the channel, two parties can purify shared entanglement with a limitation $F_{\star}<1$. 
} 
In Fig.~\ref{fig:depol}, the purification is shown up to a maximum convergent fidelity $F_{\star}$. For instance, for a channel with $p_{00}=0.75$, single rounds can increase the fidelity up to a limited one $F_\star \approx 0.863$.

\begin{figure*}[t]
\begin{minipage}{0.45\textwidth}
\centering
\includegraphics[width=\linewidth]{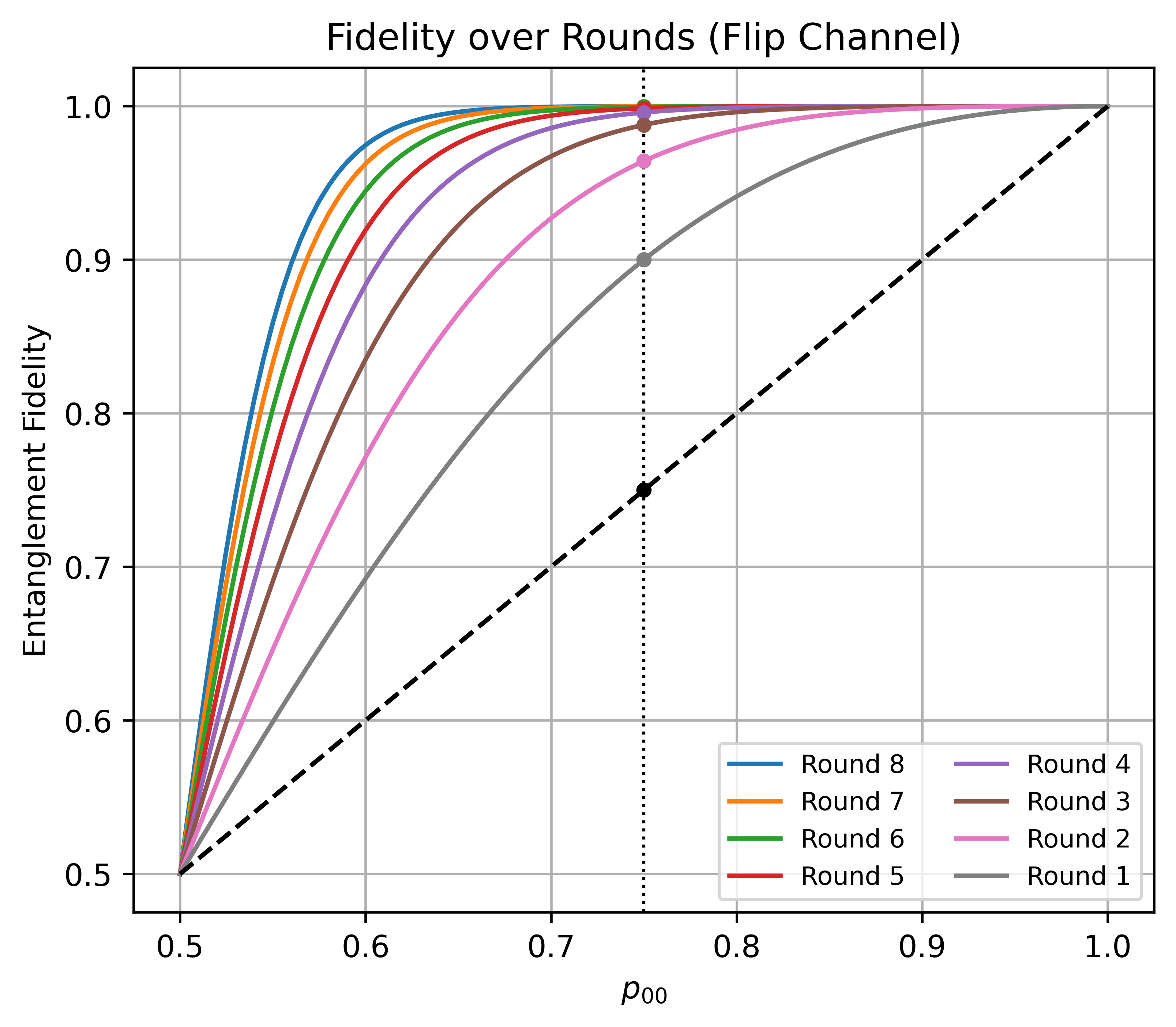}
\end{minipage}\hfill
\begin{minipage}{0.45\textwidth}
\centering
\includegraphics[width=\linewidth]{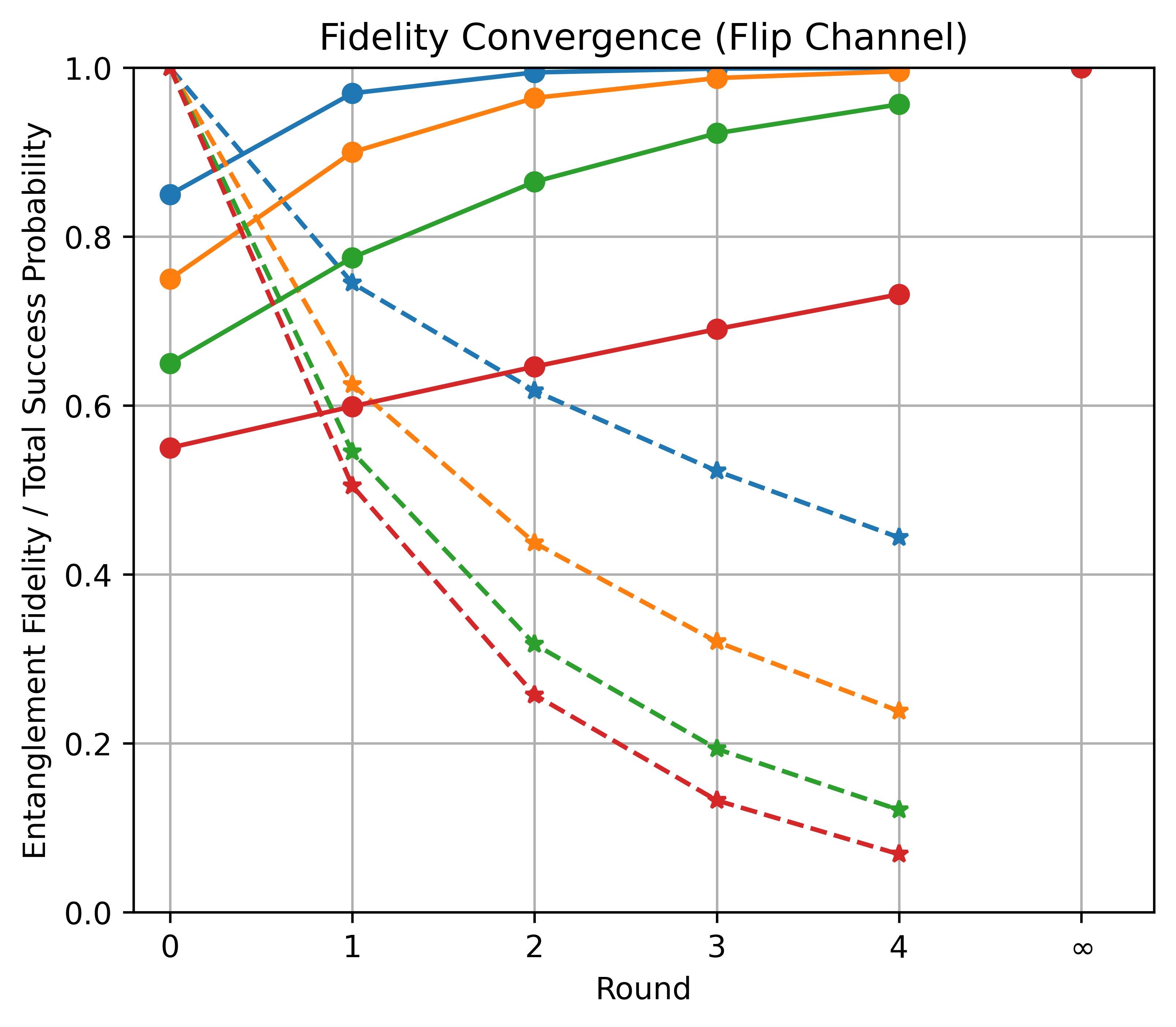}
\end{minipage}
\caption{The CAEPP with noisy transmission of a carrier against a flip channel is shown. In this case, where $Z$ errors do not occur in a channel, repeating rounds of the CAEPP achieves the purification of an ebit $F_\star=1$. 
}
\label{fig:flip}
\end{figure*}

Note that in Fig. \ref{fig:depol}, the convergent fidelity is numerically computed by repeatedly applying the relation in Eq.~\eqref{eq:coeff_update}. The total success probability for having $n$ successful rounds is given as follows,
\begin{equation}
P_{\mathrm{tot}} \;=\;  P_{\mathrm{succ}}^{(1)} \times\cdots\times P_{\mathrm{succ}}^{(r)} \times \cdots \times  P_{\mathrm{succ}}^{(n)} \label{eq:15} \end{equation}
where $P_{\mathrm{succ}}^{(r)}$ is the success probability at the $r$-th round.

The expression in Eq. (\ref{eq:15}) shows that one of $1/P_{\mathrm{tot}}$ trials is the case to achieve the $n$-round enhancement of entanglement by the CAEPP. From this, the average number of channel uses to reach the $n$-th round is computed as follows. The overall trials are classified as follows. One class contains those in which at least a single trial is successful, which are $P_{\mathrm{succ}}^{(1)}/P_{\mathrm{tot}}$. The next is cases where at least two trials are successful, which are $P_{\mathrm{succ}}^{(1)} P_{\mathrm{succ}}^{(2)}/P_{\mathrm{tot}}$. In this way, there is a class where $n$ rounds of the CAEPP are successful. Then, the number of channel uses is found as,
\bea
\frac{1}{ P_{\mathrm{tot}}} +  P_{\mathrm{succ}}^{(1)} \times \frac{1}{ P_{\mathrm{tot}} }+ \cdots +  \prod_{j=1}^{n-1} P_{\mathrm{succ}}^{(j)} \times \frac{1}{ P_{\mathrm{tot}} } \nonumber
\eea
where the first term $1/P_{\mathrm{tot}}$ is the number of channel uses in the beginning; out of them, $P_{\mathrm {succ}}^{(1)}/P_{\mathrm{tot}}$ instances are successful. In realistic scenarios, shown in Figs. \ref{fig:depol} and \ref{fig:biased_channel}, the CAEPP achieves a sufficiently high convergent fidelity for $n=4$.

Let us also consider another instance where probabilities are not equal as follows, 
\bea
\bigl(p_{00},\,\tfrac{1}{5}(1-p_{00}),\,\tfrac{2}{5}(1-p_{00}),\,\tfrac{2}{5}(1-p_{00})\bigr). \label{B_chan}
\eea
In Fig.~\ref{fig:biased_channel}, the maximum convergent fidelity by the CAEPP is shown, where \textbf{Pre-processing} applies the rotation $R_X(+\tfrac{\pi}{2})\otimes R_X(-\tfrac{\pi}{2})$ in every round, including the first, as in Eq.~\eqref{eq:coeff_update}. Compared to the example with a depolarizing channel, two parties have a higher value in the convergent fidelity. 

Note that the reason for a higher convergent fidelity, compared to a depolarizing channel, is due to the fact that the channel in \eqref{B_chan} has a lower $Z$-error rate. In fact, a Pauli channel with an even lower value of a $Z$ error can allow two parties to achieve an even higher convergent fidelity. 

\subsection{ Pauli channels for $F_{\star}=1$ } 

Let us consider Pauli channels characterized by three probabilities only. These cases can be generally parameterized as follows, 
\bea
(p_{00},\,0,\,\gamma(1-p_{00}),\,\delta(1-p_{00}))~~\mathrm{where}~~ \gamma + \delta=1.
\label{flip_chan}
\eea
Two parties exploit the channel to establish a shared state and send a carrier qubit. A shared state corresponds to the CJ state of the channel,
\bea
\rho \;=\; F\,\phi^+ + \gamma(1-F)\,\psi^+ + \delta(1-F)\,\psi^-, \nonumber
\eea
where we have used the initial fidelity $F=p_{00}$. Here, two parties omit the rotation $R_X(+\tfrac{\pi}{2})\otimes R_X(-\tfrac{\pi}{2})$ at \textbf{Pre-processing} in every round, which keeps $q_{01}=0$.

A single round of the CAEPP suppresses $X$ and $Y$ errors of the odd parity, $\psi^{\pm}$. Thus, a resulting state from the single round is a mixture of three Bell states $\phi^+$, $\psi^+$, and $\psi^-$ with a higher fidelity, 
\bea
F'\;=\; \frac{Fp_{00}}{Fp_{00} + (1-F)(1-p_{00})}.\nonumber
\eea
Hence, $X$ and $Y$ errors are suppressed. Repeating the CAEPP, we derive the fidelity after $n$ successful rounds of the protocol, 
\bea
F_n \;=\; \frac{p_{00}^{\,n+1}}{p_{00}^{\,n+1} + (1-p_{00})^{\,n+1}}, \nonumber
\eea
with the initial condition $F_0 = p_{00}$. As $n$ tends to be large, one can see that 
\bea
F_n \rightarrow 1 ~~\mathrm{as}~~n\rightarrow \infty, \nonumber 
\eea
that is, $F_\star=1$ whenever $p_{00}>\tfrac{1}{2}$.

In Fig.~\ref{fig:flip}, we consider a flip channel with probabilities $(p_{00},0,1-p_{00},0)$ that has been considered in experimental demonstrations of entanglement distillation \cite{doi:10.1126/science.aan0070, Pan:2001aa} and show that the CAEPP through noisy channels can purify noisy entanglement.

\begin{figure*}[t]
\centering
\begin{quantikz}[every arrow/.append style={no head}]
\NL & \lstick{\footnotesize $A_0$} & \gate{U_A} \QL & & & & \ctrl{2} & & & & & & & & \rstick[5]{$\rho_{\mathrm{out}}$}\\
\NL & & \midstick[2,brackets=right]{Carriers} & \lstick{$\ket{0}$} \nl & \gate{H} \QL  & \ctrl{1} & & & \permute{5} \\
\NL \lstick{$\rho$} & & \QL \nl                  & \lstick{$\ket{0}$} \nl & & \targ{} & \targ{} & \permute{5} \\
\NL & & & & & & & \gate{\mathcal{N}} & \gate{\mathcal{N}} &&& \\
\NL & \lstick{\footnotesize $B_0$} & \gate{U_B} \QL & & & & & & & \ctrl{2} & & & & & \\
\NL & &                 & & & & & & & \wire[l][1][shorten >=2.5mm]{q} & \ctrl{1} \QL & \gate{H} & \meter{0} \\
\NL & &                 & & & & & & \wire[l][1][shorten >=2.5mm]{q} & \targ{} \QL & \targ{} & & \meter{0}
\arrow[from=3-1, to=1-2, line width=0.9pt]
\arrow[from=3-1, to=5-2, line width=0.9pt]
\end{quantikz}
\caption{A successful round of the CAEPP with two carriers, using the check operators $\{X_1 X_2,\, Z_0 Z_1 Z_2\}$, is shown. Alice and Bob start from an input state $\rho$. After \textbf{Pre-processing} on a shared pair, Alice performs \textbf{Encoding} and sends the two carriers through the same noisy channel $\mathcal{N}$. Bob performs \textbf{Decoding} and measures two carriers; the shared state is kept only if the measurement outcome is $00$.}
\label{fig:two_carrier_circuit}
\end{figure*}

\subsection{ Improving the CAEPP with $F_{\star}<1$ }

It is natural to explore the possibilities of improving the CAEPP in the presence of noisy channels. 
\begin{itemize}
\item The first strategy is to exploit two-way entanglement distillation in Refs. \cite{bennett1996purification, Deutsch1996}. Once a shared state reaches the maximum convergent fidelity, copies of them are collected and exploited in two-way EPPs. 
\item The second strategy is to enhance the CAEPP by leveraging additional resources.
\end{itemize}
In what follows, we present the CAEPP with multiple qubits as carriers against noisy channels and show that the protocol leads to pure entanglement.

\begin{figure}[t]
\centering
\begin{minipage}{\columnwidth}
\raggedright
\algcaption{mCAEPP}{CAEPP}
\begin{algorithmic}[1]
\State \textbf{Initialization:} Alice and Bob share a noisy entangled pair \label{alg:init}
\Repeat \Comment{One round = Steps~\ref{alg:manipulate_state}--\ref{alg:end}}
    \State \textbf{Pre-processing:} Apply LOCC operations to the shared pair\label{alg:manipulate_state}
    \State \textbf{Encoding (Alice):} Prepare $m$ carrier qubits in $\ket{0}^{\otimes m}$, apply an encoding operation, and send them through the channel \label{alg:Acar}
    \State \textbf{Decoding (Bob):} Apply a decoding operation and measure the carriers in $Z$ basis\label{alg:Bcar}
    \State \textbf{Communication:} Bob sends $m$ syndrome bits to Alice \label{alg:comm}
    \State \textbf{Decision:} \label{alg:decision}
    \If{the syndrome bits are all $0$}
        \State \textbf{Success:} Keep the shared pair \label{alg:succ}
    \Else
        \State \textbf{Failure:} Discard and restart from Step~\ref{alg:init}
    \EndIf \label{alg:end}
\Until{the target fidelity is achieved}
\end{algorithmic}
\algfinish
\end{minipage}
\end{figure}

\section{CAEPP with multiple carriers} \label{sec:multiple_carrier}

We generalize the CAEPP with multiple qubits, called \emph{multi-carrier-assisted EPP (mCAEPP)}, which purifies a single-copy shared state to an ebit through noisy channels. The protocol is summarized in Algorithm~\ref{CAEPP}.



{
\subsection{Pauli check operators}
\paragraph*{Preliminaries}
The purification protocol with multiple carriers can be described in the stabilizer formalism \cite{Dur2003MultiparticleGraphStates, Matsumoto2003Conversion, GlancyKnillVasconcelos2006}.
Here, we use the stabilizer formalism primarily as a Pauli check measurement framework.
A \emph{Pauli check operator} is a Hermitian Pauli observable $S\in\mathcal{P}_n$ with eigenvalues $\pm1$, where $\mathcal{P}_{n}:=\{I,X,Y,Z\}^{\otimes(n)}$ (up to phases) denotes the $n$-qubit Pauli group.
A commuting family of checks $\{S_i\}$ can be measured jointly, and postselecting on outcomes $+1$ projects onto their simultaneous $(+1)$ eigenspace.
For a Pauli error $E$, the syndrome bit of $S_i$ is determined by commutation: it flips ($-1$) iff $E$ anticommutes with $S_i$.

We consider the $(m+2)$-qubit register $\tilde{\mathsf{R}}:=(A,B,1,\ldots,m)$,
where $A$ and $B$ denote the two memory qubits of the shared pair, and $1,\ldots,m$ label the carrier qubits. 
In each round of mCAEPP, we choose a commuting set of $m$ check operators $\{\tilde S_i\}_{i=1}^m\subset\mathcal{P}_{m+2}$ on $\tilde{\mathsf{R}}$.
We choose $m$ commuting checks because the protocol extracts exactly $m$ syndrome bits from the 
$m$ carrier measurements.
Alice applies a \textit{encoding} operation that correlates the carriers with the shared pair according to these checks, sends the $m$ carriers through the Pauli channel $\mathcal{N}$, and Bob applies a \textit{decoding} operation and measures the carriers in the $Z$ basis. 
The measurement outcomes provide the \emph{syndrome bits}: we record $0$ for eigenvalue $+1$ and $1$ for eigenvalue $-1$, and we accept the round if and only if all syndrome bits are $0$.

\paragraph*{Effective register}
To implement syndrome extraction by carrier measurements, it is convenient to work with an effective description in which Pauli errors on $A$ are relocated to $B$ using the transpose property of $\ket{\phi^{+}}_{AB}$,
\begin{equation*}
(M\otimes \I)\ket{\phi^{+}} = (\I\otimes M^\top)\ket{\phi^{+}}.
\end{equation*}
Accordingly, we introduce effective Pauli check operators $\{S_i\}_{i=1}^m$ acting on the $(m+1)$-qubit subsystem $\mathsf R = (0,1,\ldots,m)$, where the effective qubit $0$ represents subsystem $B$ after relocating errors from $A$ to $B$. Each $S_i$ reproduces the same syndrome statistics as the corresponding physical check $\tilde S_i$ acting on $\tilde{\mathsf{R}}=(A,B,1,\ldots,m)$. 
For instance, on $\ket{\phi^{+}}_{AB}$ the physical check $Z_A Z_B Z_1\cdots Z_m$ is equivalent to the effective check $Z_0 Z_1\cdots Z_m$. 
In the remainder of the paper, unless stated otherwise, ``check operators'' refer to the effective checks $\{S_i\}$ acting on the effective register $\mathsf R=(0,1,\ldots,m)$. Note that despite the reduced notation, the physical protocol is implemented on the full register $\tilde{\mathsf{R}}$.

\subsection{Syndrome extraction}

\paragraph*{Encoding and decoding operations}
Fix a commuting set of $m$ Pauli check operators $\{S_i\}_{i=1}^m\subset\mathcal{P}_{m+1}$ to be extracted in a given round. The purpose of the encoding/decoding operations is to realize projective measurements of these checks using $Z$-basis measurements on the carriers. Concretely, we choose an encoding operation $U_{\mathrm{enc}}$ such that
\begin{equation}
U_{\mathrm{enc}}\, Z_i \,U_{\mathrm{enc}}\dg = S_i,
\qquad \forall i\in\{1,\ldots,m\}, \label{eq:conjugation}
\end{equation}
i.e., $U_{\mathrm{enc}}$ maps the trivial carrier stabilizer $\langle Z_1,\ldots,Z_m\rangle$ to the intended set of checks $\{S_i\}$ on $\mathsf{R}$. 

After transmission of the carriers through the Pauli channel $\mathcal{N}$, Bob obtains the syndrome by applying a decoding operation $U_{\mathrm{dec}}$ and measuring each carrier in the $Z$ basis. A convenient choice is $U_{\mathrm{dec}}=U_{\mathrm{enc}}\dg$, so that
\(
U_{\mathrm{dec}}\dg\, Z_i\, U_{\mathrm{dec}}
= U_{\mathrm{enc}}\, Z_i\, U_{\mathrm{enc}}\dg
= S_i.
\)
Equivalently, measuring $Z_i$ on carriers after applying $U_{\mathrm{dec}} = U_{\mathrm{enc}}\dg$ implements a measurement of $S_i$ on the pre-decoding state.

\paragraph*{Detected errors}
Let $E$ be a Pauli error acting on $\mathsf{R}$. The extracted syndrome bits are determined by the commutation relations between $E$ and the check operators: $E$ is \emph{detected} (yields a nontrivial syndrome) if it anticommutes with at least one $S_i$, i.e., $E S_i = - S_i E$ for some $i$, whereas errors commuting with all $\{S_i\}$ pass the checks and remain undetected.
Thus the protocol postselects onto the simultaneous $+1$ eigenspace of $\{S_i\}$ by accepting iff the measured syndrome is all zeros.

\subsection{Two-carrier instance}
Fig.~\ref{fig:two_carrier_circuit} illustrates a successful round of the CAEPP with two carriers using the effective check operators
\begin{align}
\{X_1 X_2,\; Z_0 Z_1 Z_2\}. \label{stab2}
\end{align}
A successful round means that Bob has syndrome bits $00$, both check operators measured $+1$. A Pauli error $E$ is \emph{detected} precisely when it anticommutes with at least one check operator.

\subsubsection*{Check operator $X_1 X_2$}
This detects single-qubit errors $Z$ or $Y$ on a carrier, e.g., $Z_1$, $Z_2$, $Y_1$, $Y_2$, since $X$ anticommutes with $Z$ and $Y$ on the same qubit. By contrast, $Z_1 Z_2$ commutes with $X_1 X_2$ and is not detected by this check operator.

\subsubsection*{Check operator $Z_0 Z_1 Z_2$}
This detects errors with an \emph{odd} number of $X$ and $Y$ errors on the triplet $\{0,1,2\}$, e.g., $X_0$, $Y_1$, $X_2$, $X_0 X_1 X_2$, $X_0 Y_1 X_2$.
}

\begin{figure*}[t]
\begin{minipage}{0.45\textwidth}
\centering
\includegraphics[width=\linewidth]{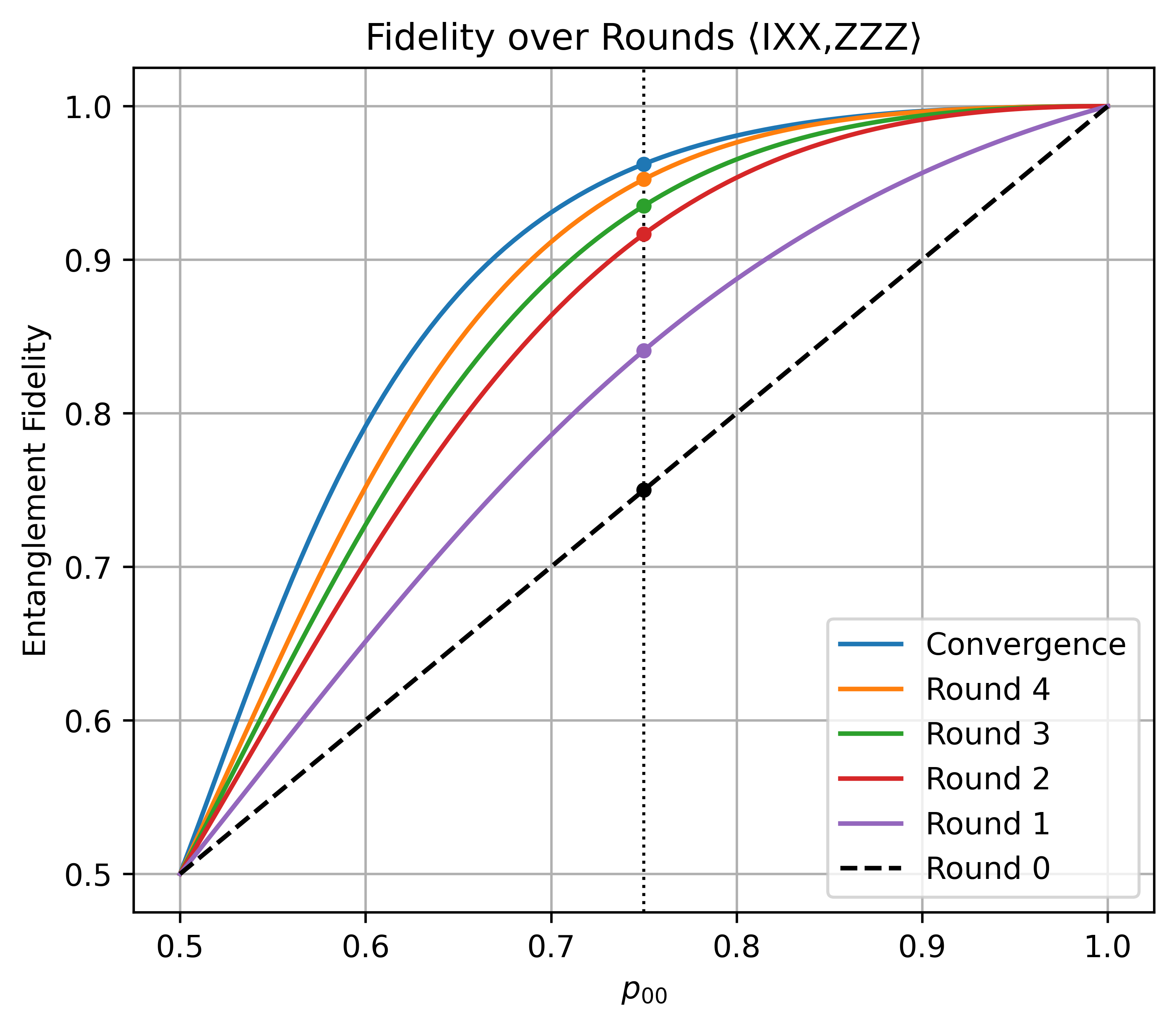}
\end{minipage}\hfill
\begin{minipage}{0.45\textwidth}
\centering
\includegraphics[width=\linewidth]{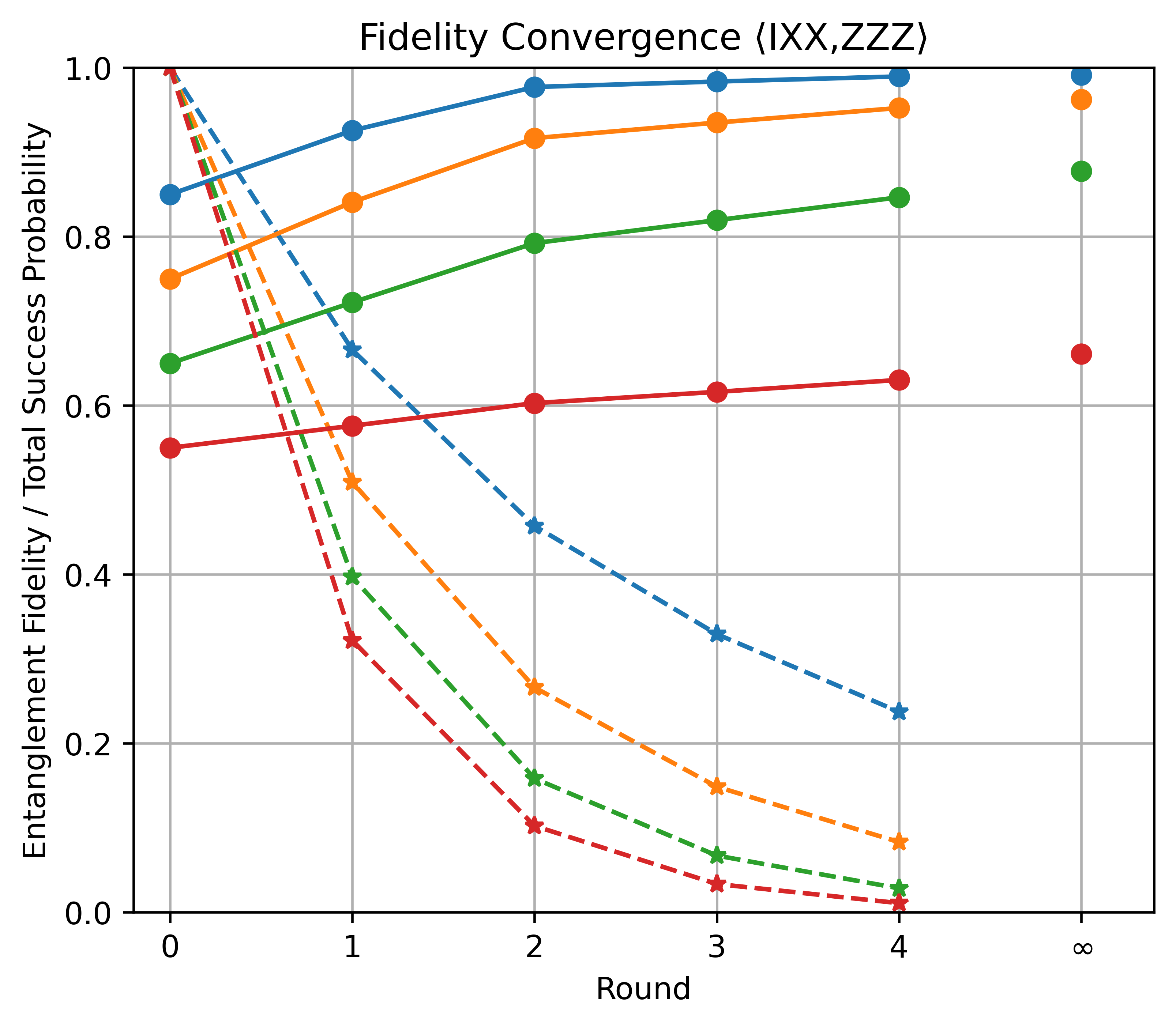}
\end{minipage}
\caption{  { (Left) The fidelity increases by the mCAEPP with two carriers. Compared to the CAEPP with a single carrier in Fig. \ref{fig:depol}, the maximum convergent fidelity is even higher. For instance, for $F=p_{00}=0.75$, the maximum fidelity is over $0.95$; with a single carrier, it is $0.86$. (Right) A total success probability is computed as more rounds are executed. Four rounds increase the entanglement fidelity sufficiently close to the convergent one; the probability is also shown according to an initial fidelity. For instance, starting with a fidelity $0.75$ (orange), the total probability is about $0.1$.} }
\label{fig:two_carrier_graph}
\end{figure*}

As it is shown in Fig.~\ref{fig:two_carrier_graph}, under a depolarizing channel the two-carrier scheme achieves a strictly higher fidelity in every round and a higher fidelity bound than the single-carrier case in Fig.~\ref{fig:depol}.

\begin{figure*}[t]
\begin{minipage}{0.31\textwidth}
\centering
\includegraphics[width=\textwidth]{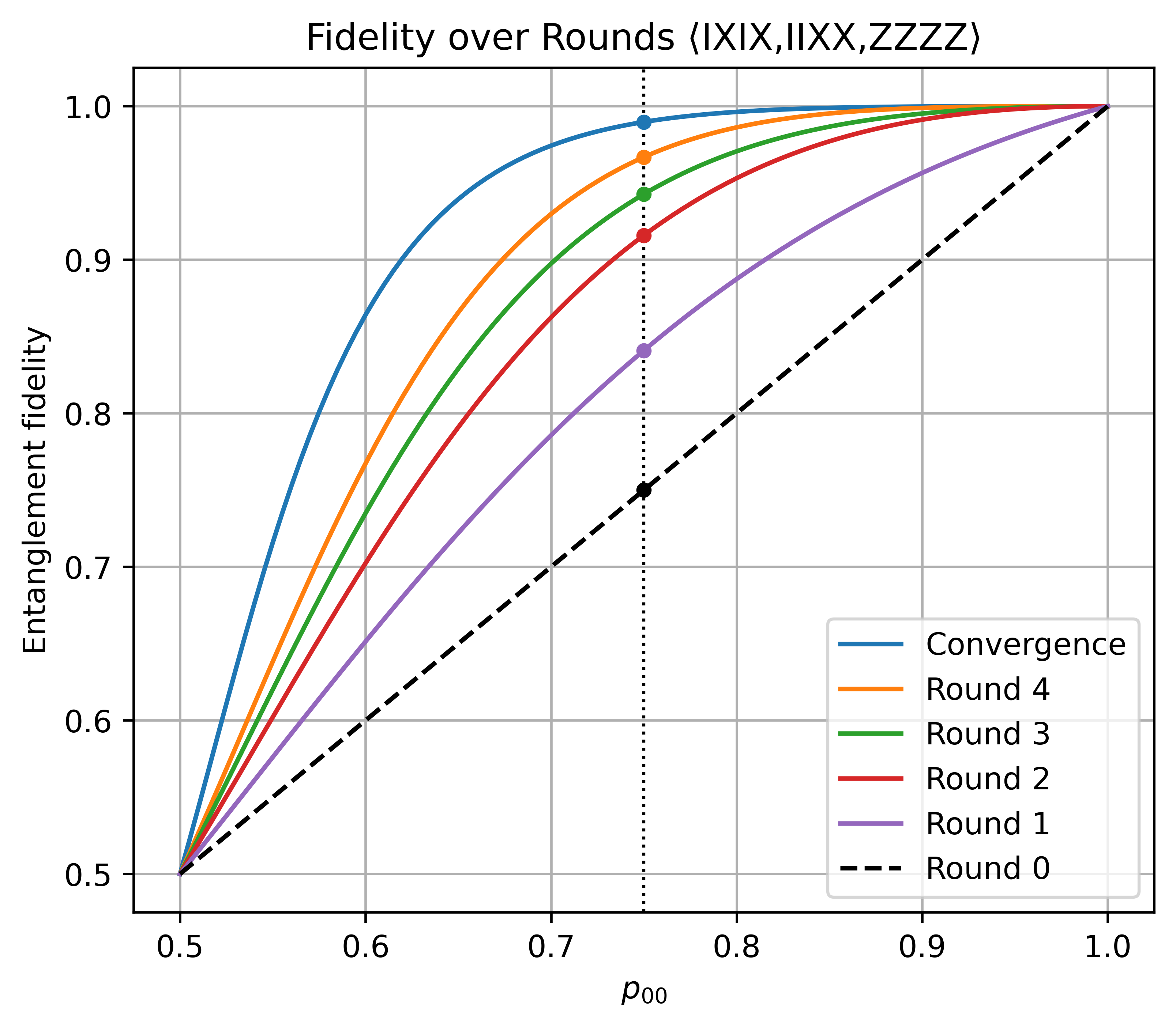}
\newline (a)
\end{minipage}
\hfill
\begin{minipage}{0.31\textwidth}
\centering
\includegraphics[width=\textwidth]{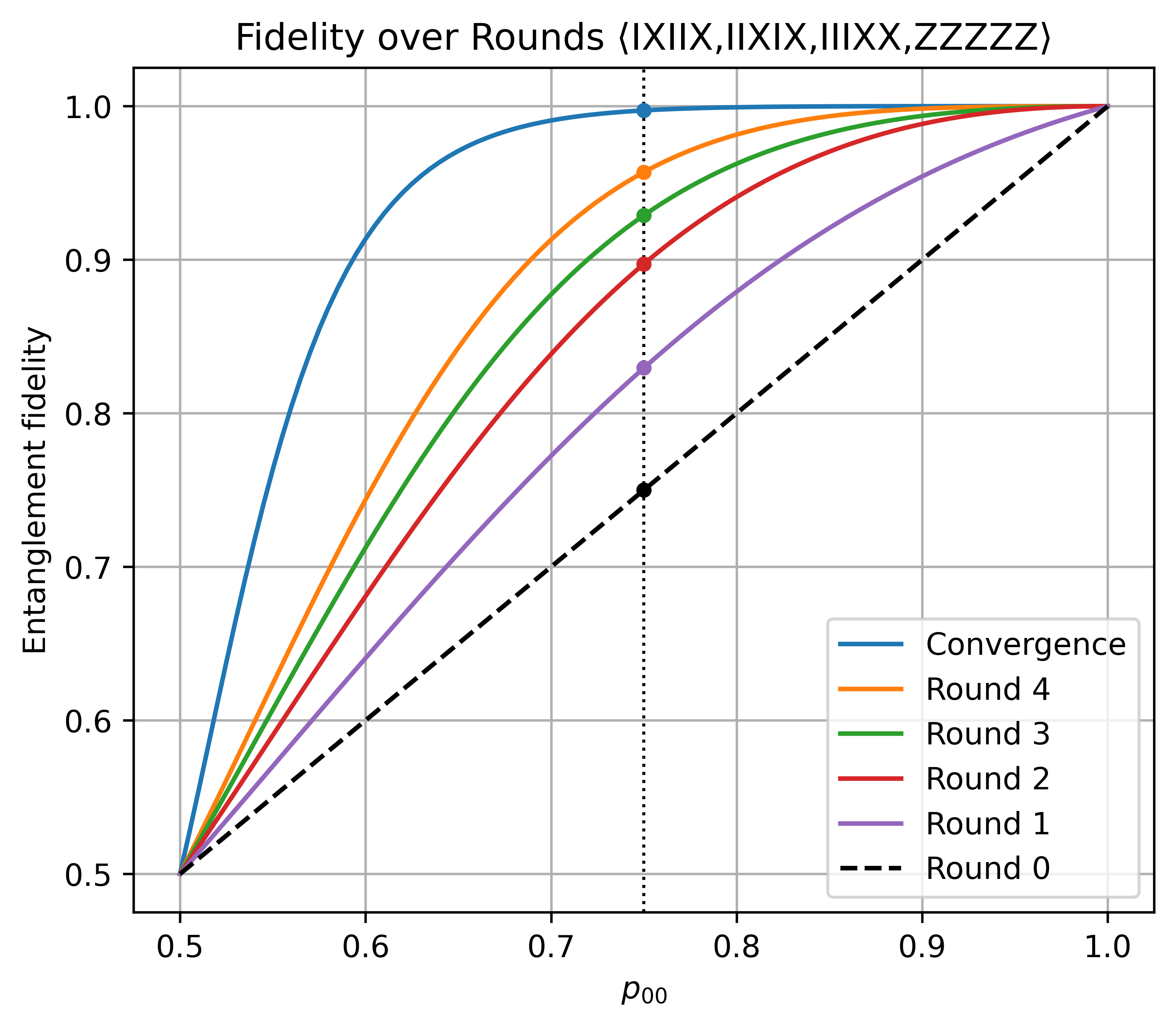}
\newline (b)
\end{minipage}
\hfill
\begin{minipage}{0.31\textwidth}
\centering
\includegraphics[width=\textwidth]{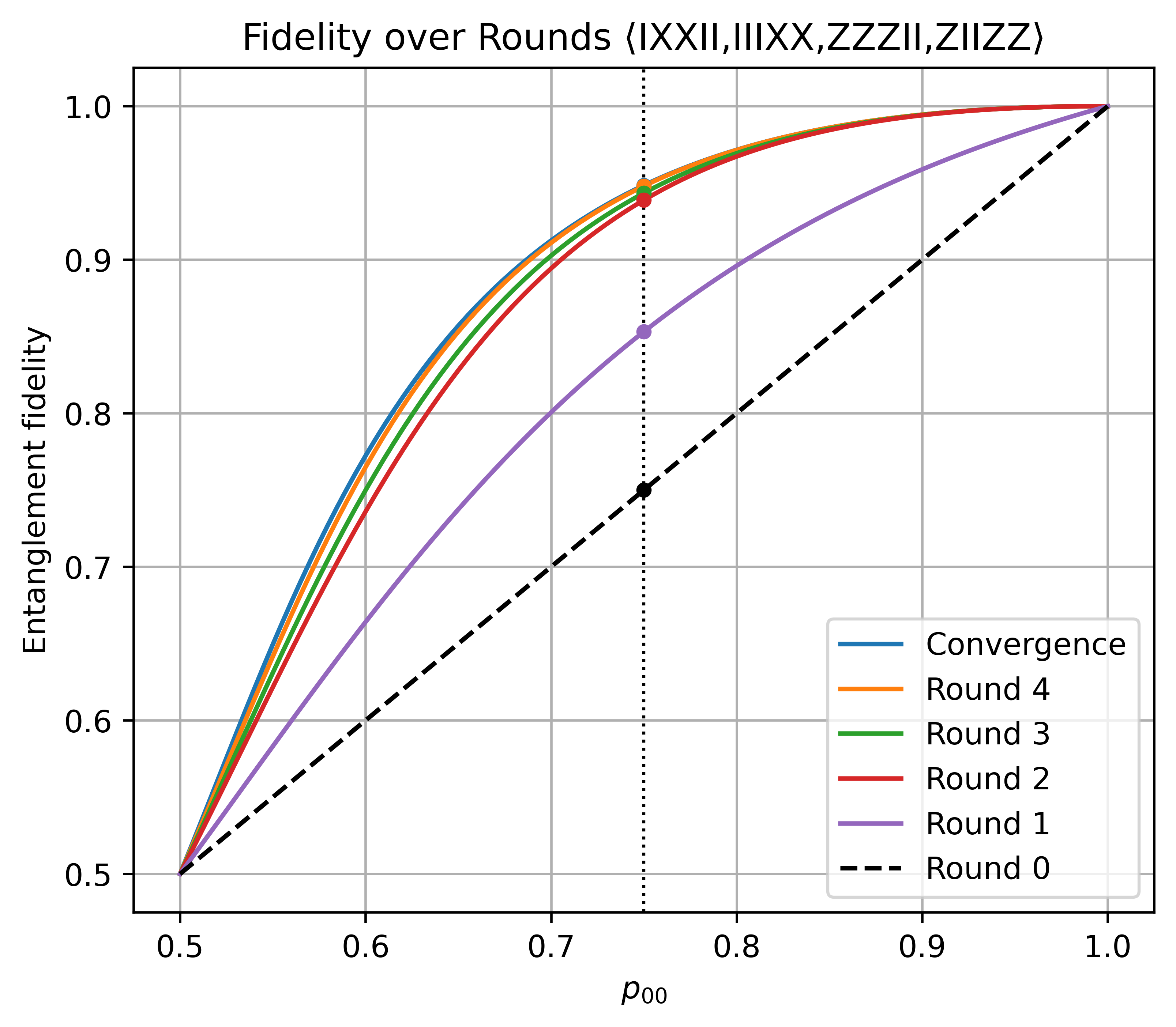}
\newline (c)
\end{minipage}
\caption{The mCAEPP with various choices of stabilizers is shown. (a) Three carriers with ``star'' check operators $\{X_1 X_3,\, X_2 X_3,\, Z_0 Z_1 Z_2 Z_3\}$. 
(b) Four carriers with ``star'' check operators $\{X_1 X_4,\, X_2 X_4,\, X_3 X_4,\, Z_0 Z_1 Z_2 Z_3 Z_4\}$. 
(c) Four carriers with ``pairwise'' check operators $\{X_1 X_2,\, Z_0 Z_1 Z_2,\, X_3 X_4,\, Z_0 Z_3 Z_4\}$. }
\label{fig:three_four_carriers}
\end{figure*}

\subsection{Beyond two carriers: $F_{\star}=1$}
\label{sec:fidelity_to_one}

The maximum convergent fidelity increases with a larger number of carrier qubits, and the enhancement relies on the selection of stabilizers. In Fig.~\ref{fig:three_four_carriers}, various check operators are exploited to show higher maximum convergent fidelities.

\begin{figure}[t]
\centering
\includegraphics[width=0.95\linewidth]{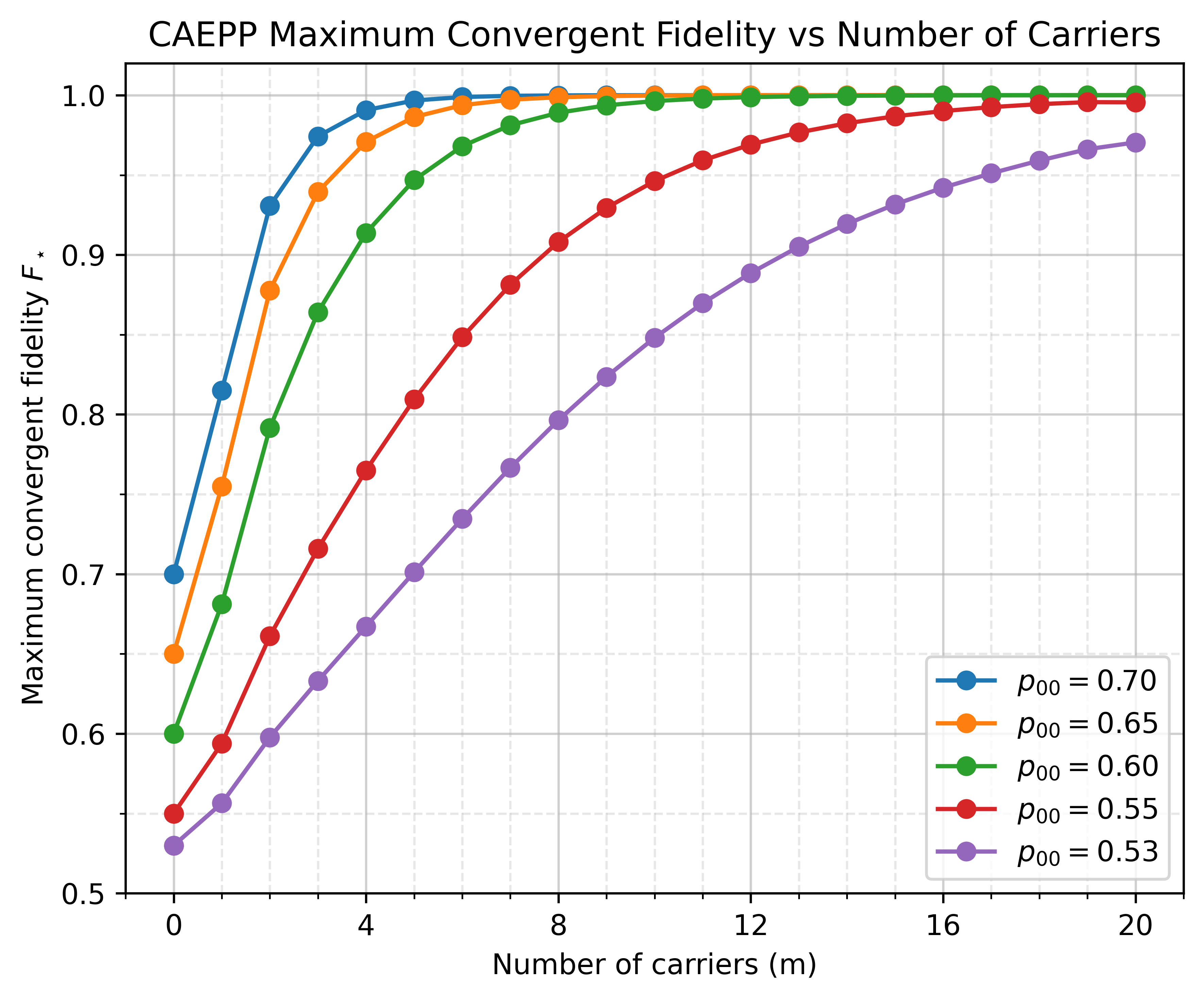}
\caption{For the CAEPP against a depolarizing channel, the maximum convergent bound $F_\star$ improves by a larger number of carriers $m$. Channels with fidelities $p_{00}\in\{0.53,\,0.55,\,0.60,\,0.65,\,0.70\}$ are considered. For all cases, the maximum fidelity converges to $1$ as $m$ is sufficiently large.}
\label{fig:fid_bound_m}
\end{figure}

In fact, the mCAEPP generally purifies noisy entanglement with a sufficiently large number of carriers, so that $F_\star \to 1$, for any Pauli channel with $p_{00} > \tfrac{1}{2}$. {We recall that, as shown in Section \ref{sec:preliminaries}, a Pauli channel is entanglement-breaking if and only if $p_{00} \leq 1/2$, see also Eq. (\ref{eq:order}), in which single-qubit transmission can be simulated by LOCC. Two parties can apply carrier encoding and decoding against a noisy Pauli channel, with the random-unitary strategy with unitary $2$-design, shown in Section~\ref{sec:channel_twirling}. In this case, two parties share a depolarizing channel \cite{PhysRevA.99.062302, 10993387}.}

{Then, under a Pauli channel with $p_{00}>\tfrac{1}{2}$, using the ``star'' check operators for $m$ carriers,
\begin{align}
\{X_1 X_m,\, X_2 X_m,\,\ldots,\,X_{m-1} X_m,\, Z_0 Z_1 \cdots Z_m\},
\end{align}
the achievable fidelity bound $F_\star$ converges to $1$ as $m\to\infty$. A detailed proof is given in Appendix~\ref{App}.
Fig.~\ref{fig:fid_bound_m} illustrates that $F_\star$ approaches $1$ as $m$ grows and, numerically, increases monotonically with $m$. 
While $F_\star$ increases with $m$, the success probability per round decreases with $m$; thus $m$ should be chosen to balance target fidelity and resources.}

{
\subsection{ Collective entanglement pumping}
Let us recall that entanglement pumping with a fixed elementary pair, denoted by $\sigma$ that is not maximally entangled, cannot distill a maximally entangled state \cite{Dur:2007aa}. Depending on how entangled an elementary pair is, entanglement pumping can achieve a convergent {fidelity} which is strictly away from being maximally entangled. This limitation is also reflected in the CAEPP with noisy single-qubit carriers, as discussed in Section~\ref{sec:equiv}. 


Our result that mCAEPP can distill a maximally entangled state applies to devising a \emph{collective} pumping step in which multiple copies $\sigma^{\otimes m}$ are used as an elementary pair. Hence, the mCAEPP resolves the limitation of entanglement pumping.
}




{
\section{Comparison: the CAEPP versus TWEPP} \label{sec:comparison}
We here review two-way entanglement purification protocols (TWEPPs) with multiple copies via a stabilizer code~\cite{Matsumoto2003Conversion, GlancyKnillVasconcelos2006}, and consider the examples of two-pair protocols~\cite{bennett1996purification, Deutsch1996}. We then compare the CAEPP and TWEPPs in terms of resources for their practical realizations. 
}

\subsection{TWEPPs}

At \textbf{Initialization}, Alice and Bob share copies of entangled pairs $\rho^{\otimes n}$ for large $n$. A TWEPP contains steps as follows. 
\begin{enumerate}
\item \textbf{Selection:} Choose $m+1$ noisy pairs.
\item \textbf{Pre-processing:} Apply a pre-designed LOCC to the pairs.
\item \textbf{Decoding (Alice):} Apply a stabilizer decoding operation on Alice’s $m+1$ qubits and measure $m$ of them (keeping the first).
\item \textbf{Decoding (Bob):} Apply the conjugate decoding operation on Bob’s $m+1$ qubits and measure $m$ of them.
\item \textbf{Communication:} Exchange their $m$-bit syndrome strings.
\item \textbf{Decision:} If the syndrome strings from Alice and Bob match, declare \textbf{Success} and keep the first pair; otherwise, declare \textbf{Failure} and discard it.
\end{enumerate}
Rounds repeat until a pair reaches the target fidelity.

\begin{figure}[t]
\centering
\begin{minipage}{\columnwidth}
\raggedright
\algcaption{Two-way EPP}{TWEPP}
\begin{algorithmic}[1]
\State \textbf{Initialization:} Alice and Bob share a sufficiently large number of noisy entangled pairs \label{Alg:init}
\Repeat \Comment{One round = Steps~\ref{Alg:select}--\ref{Alg:end}}
    \State \textbf{Input selection:} Select $m+1$ pairs from the shared pool \label{Alg:select}
    \State \textbf{Pre-processing:} Apply LOCC preprocessing to the selected pairs \label{Alg:pre}
    \State \textbf{Decoding (Alice):} Apply a decoding operation and measure $m$ qubits (keep the first) \label{Alg:Adec}
    \State \textbf{Decoding (Bob):} Apply the conjugate decoding operation and measure $m$ qubits (keep the first) \label{Alg:Bdec}
    \State \textbf{Communication:} Exchange syndrome strings \label{Alg:comm}
    \State \textbf{Decision:}\label{Alg:decision}
    \If{syndrome strings match}
        \State \textbf{Success:} Keep the remaining pair \label{Alg:succ}
    \Else
        \State \textbf{Failure:} Discard the pair
    \EndIf \label{Alg:end}
\Until{a pair achieves the target fidelity}
\end{algorithmic}
\algfinish
\end{minipage}
\end{figure}

\subsubsection*{Examples}
For $m=1$, it {suffices} to use the check operator $ZZ$, i.e., a CNOT gate from the first pair (control) to the second pair (target) at the \textbf{Decoding} steps. The two common \textbf{Pre-processing} strategies are:
\begin{itemize}
\item \textbf{BBPSSW \cite{bennett1996purification}:} { Twirling operations transform unknown states to copies of isotropic states that have coefficients $q_{01}=q_{10}=q_{11}=(1-q_{00})/3$ \cite{PhysRevA.60.1888}.}
\item \textbf{DEJMPS \cite{Deutsch1996}:} Applying $R_X(+\tfrac{\pi}{2})\otimes R_X(-\tfrac{\pi}{2})$, which swaps $\phi^-\!\leftrightarrow\!\psi^-$, yielding $q_{00} \geq q_{11}\geq q_{10}\ge q_{01}$ and reducing the $\phi^-$ ($Z$-error) weight, thereby achieving higher success-conditioned fidelity.
\end{itemize}

\begin{figure}[t]
\centering
\begin{quantikz}[every arrow/.append style={line width=0.9pt, no head}]
& \nl & \nl \lstick{\footnotesize $A_0$} & \gate{U_A} & \ctrl{1}& & & \rstick[4]{$\rho_{\mathrm{out}}$}\\
\lstick{$\rho$} & \nl & \nl \lstick{\footnotesize $A_1$} & \gate{U_A} & \targ{} & \meter{m} \\
\lstick{$\sigma$} & \nl & \nl \lstick{\footnotesize $B_0$} & \gate{U_B} & \ctrl{1}& & & \\
& \nl & \nl \lstick{\footnotesize $B_1$} & \gate{U_B} & \targ{} & \meter{m} \\
\arrow[from=2-1, to=1-3]
\arrow[from=2-1, to=3-3]
\arrow[from=3-1, to=2-3]
\arrow[from=3-1, to=4-3]
\end{quantikz}
\caption{A successful round of the two-pair TWEPP. Two input states $\rho$ and $\sigma$ are shown. After preprocessing, Alice and Bob apply local CNOTs and measure the second pair. After exchanging measurement outcomes, they keep the first pair if and only if their outcomes agree.}
\label{fig:two_pair}
\end{figure}


Fig.~\ref{fig:two_pair} illustrates a purification round of two-pair TWEPP.

\subsubsection*{Realistic constraints}
TWEPPs face several drawbacks in realistic scenarios:
\begin{enumerate}
\item \textbf{Identical input requirement.} A universal EPP that always guarantees an output fidelity no worse than that of each input state cannot be implemented using LOCC~\cite{Zang2025NoGo}. Consequently, a TWEPP may reduce entanglement depending on the inputs. To avoid this, TWEPPs typically assume identical input pairs.
\item \textbf{Quantum-memory demand.} TWEPPs require quantum memory for two reasons: (i) short-term memory to synchronize two qubits for bilateral CNOT gates, and (ii) long-term memory to store successful output pairs until an identical pair is available for the subsequent round. Both types of quantum memory remain technically challenging. Moreover, TWEPPs demand large-capacity quantum memories to hold multiple pairs during the process. Recent work has aimed to reduce this memory cost~\cite{Zheng2022Entanglement, Davies2024Entanglement}.
\item \textbf{Sensitivity to measurement noise.} Measurement noise reduces the entanglement fidelity after each purification round and can even prevent entanglement distillation from weakly entangled states~\cite{Dur1999QuantumRepeaters}.
\end{enumerate}

\begin{table*}[t]
\caption{Operational distinctions between the CAEPP and TWEPP in stabilizer-based schemes (per round)}
\label{tab:caepp-vs-twepp}
\centering
\renewcommand{\arraystretch}{1.2}
\begin{tabular}{p{0.32\linewidth} p{0.32\linewidth} p{0.32\linewidth}}
\hline\hline
\multicolumn{1}{c}{\textbf{Aspect}} & \multicolumn{1}{c}{\textbf{CAEPP} ($m$ carriers)} & \multicolumn{1}{c}{\textbf{TWEPP} ($m+1$ pairs)} \\
\hline
memory demand & one stored pair (2 qubits) & $m+1$ stored pairs ($2(m+1)$ qubits) \\
input selection flexibility & fixed noisy channel & flexible (typically identical pairs) \\
encoding / decoding & encoding (Alice) + decoding (Bob) & decoding only (Alice and Bob) \\
quantum communication & $m$ carrier transmissions & none (pre-shared pairs only) \\
measurements & $m$ carrier qubits (Bob) & $m$ qubits per party (Alice and Bob) \\
classical communication & $m$ one-way syndrome bits (Bob $\to$ Alice) & $m$ two-way syndrome bits (Alice $\leftrightarrow$ Bob) \\
\hline\hline
\end{tabular}
\end{table*}

{

\subsection{ Equivalent performance of CAEPP and EP} \label{sec:equiv}

Consider a single round in the entanglement pumping (EP). Suppose that Alice and Bob share a Bell-diagonal state $\rho$ in a long-term quantum memory, and the elementary pair is given by $\sigma$:
\begin{align}
\rho &= q_{00}\phi^+ + q_{01}\phi^- + q_{10}\psi^+ + q_{11}\psi^-, \nonumber \\
\sigma &= p_{00}\phi^+ + p_{01}\phi^- + p_{10}\psi^+ + p_{11}\psi^-. \label{eq:sigmarho}
\end{align}
A successful round (the parity check with bilateral CNOT gates followed by measurements, preceded by \textbf{Pre-processing} $R_X(+\tfrac{\pi}{2})\otimes R_X(-\tfrac{\pi}{2})$ on $\rho$) updates the shared state to $\rho' = q'_{00}\phi^+ + q'_{01}\phi^- + q'_{10}\psi^+ + q'_{11}\psi^-$, where 
\begin{align}
\bm{q'} &= \frac{1 }{P_\mathrm{succ}}L \bm q, 
\quad 
L = 
\begin{bmatrix}
p_{00} & 0 & 0 & p_{01}\\
p_{01} & 0 & 0 & p_{00}\\
0 & p_{11} & p_{10} & 0\\
0 & p_{10} & p_{11} & 0
\end{bmatrix}, \label{eq:coeff_updateep}
\end{align}
where $P_\mathrm{succ} = \bm{1}^\top L \bm q$ denotes the success probability and $\bm{1} = (1,1,1,1)^\top$. We recall that the resulting fidelities in Eq. (\ref{eq:coeff_updateep}) are identical to those by a single round of the CAEPP in Eq. (\ref{eq:coeff_update}); hence, it is proven that both the CAEPP and EP equally enhance entanglement, where the success probabilities are also identical. Note also a technical explanation that, through the CJ isomorphism, an elementary pair in EP corresponds to a CJ state of the carrier-transmission channel. 
}

\tikzset{
  cnotHL/.style={rounded corners, fill=red!15, draw=red!40, inner xsep=1pt, inner ysep=0pt},
  measHL/.style={fill=yellow!25, draw=black},
  memHL/.style={fill=violet!20, draw=black},
}
\begin{figure*}[t]
\centering
\begin{minipage}{0.48\textwidth}
\centering
\begin{quantikz}[every arrow/.append style={no head}]
\nl & \nl \lstick{\footnotesize $A_0$}
& \gate[style={memHL}]{\mathrm{QM}} & \ctrl{1}\gategroup[2,steps=1,style={cnotHL},background]{} & & & & \rstick[5]{$\rho_{\mathrm{out}}$}\\
\nl & \nl & \nl \midstick{$\ket{0}$} & \targ{} & \permute{3}  \\
\lstick{$\rho_{\mathrm{iso}}$} \NL &&&& \gate{\mathcal{N}} &&& \\
\NL & & & & & \targ{}\gategroup[2,steps=1,style={cnotHL},background]{} \wire[l][1][shorten >=2.5mm]{q} & \meter[style={measHL}]{0} \QL \\
\nl & \nl \lstick{\footnotesize $B_0$} & \gate[style={memHL}]{\mathrm{QM}} & & & \ctrl{-1} & &
\arrow[from=3-1, to=1-2, line width=0.9pt]
\arrow[from=3-1, to=5-2, line width=0.9pt]
\end{quantikz}
\newline (a)
\end{minipage}
\hfill
\begin{minipage}{0.48\textwidth}
\centering
\begin{quantikz}[every arrow/.append style={line width=0.9pt, no head}]
& \nl & \nl \lstick{\footnotesize $A_0$}
& \gate[style={memHL}]{\mathrm{QM}}
& \ctrl{1}\gategroup[2,steps=1,style={cnotHL},background]{}
& & & \rstick[3]{$\rho_{\mathrm{out}}$}\\
\lstick{$\rho_{\mathrm{iso}}$} & \nl & \nl \lstick{\footnotesize $A_1$}
& \gate[style={memHL}]{\mathrm{QM}}
& \targ{}
& \meter[style={measHL}]{m} \\
\lstick{$\rho_{\mathrm{iso}}$} & \nl & \nl \lstick{\footnotesize $B_0$}
& \gate[style={memHL}]{\mathrm{QM}}
& \ctrl{1}\gategroup[2,steps=1,style={cnotHL},background]{}
& & & \\
& \nl & \nl \lstick{\footnotesize $B_1$}
& \gate[style={memHL}]{\mathrm{QM}}
& \targ{}
& \meter[style={measHL}]{m}
\arrow[from=2-1, to=1-3]
\arrow[from=2-1, to=3-3]
\arrow[from=3-1, to=2-3]
\arrow[from=3-1, to=4-3]
\end{quantikz}
\newline \rule{0pt}{2.5em} (b) \rule{0pt}{2.5em}
\label{fig:noise-twepp-overwrite}
\end{minipage}
\caption{ The effect of noise in the CAEPP and TWEPPs is compared. (a) The CAEPP has two quantum memories ($\mathrm{QM}$) (violet), two CNOT gates (red), and a single-qubit measurement (yellow), all of which can be noisy. (b) TWEPP needs four quantum memories ($\mathrm{QM}$) (violet), two CNOT gates (red), and two single-qubit measurements (yellow). Noisy quantum memories and noisy single-qubit measurements affect TWEPPs more than the CAEPP.}
\label{fig:noise_sources}
\end{figure*}

\begin{figure*}[t]
\centering
\begin{minipage}{0.32\textwidth}
\centering
\includegraphics[width=\linewidth]{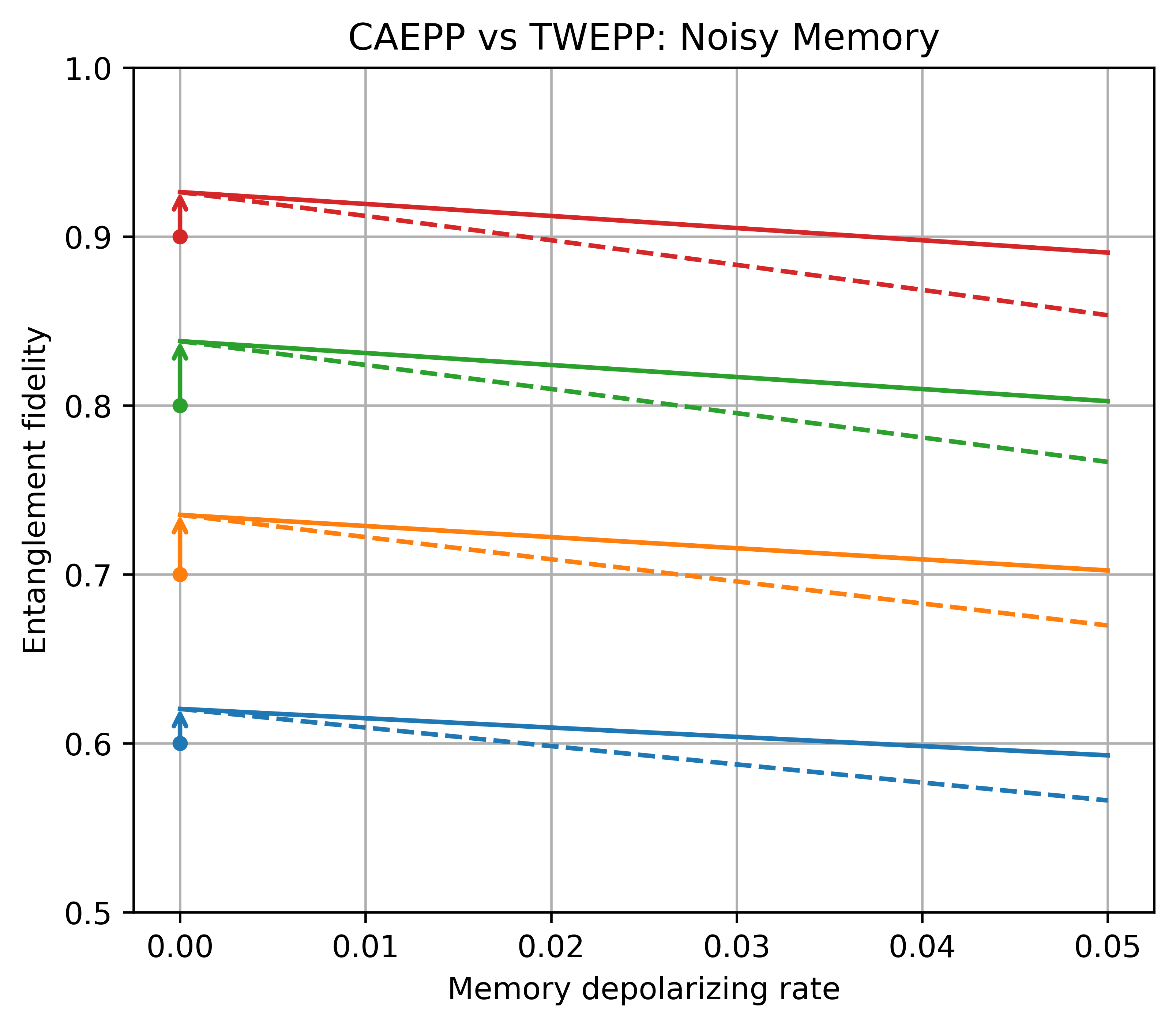}
\subcaption{
The entanglement fidelity after a single round \textbf{Success} decreases by a memory error with a rate $e$.}
\label{fig:memory_error}
\end{minipage}
\hfill
\begin{minipage}{0.32\textwidth}
\centering
\includegraphics[width=\linewidth]{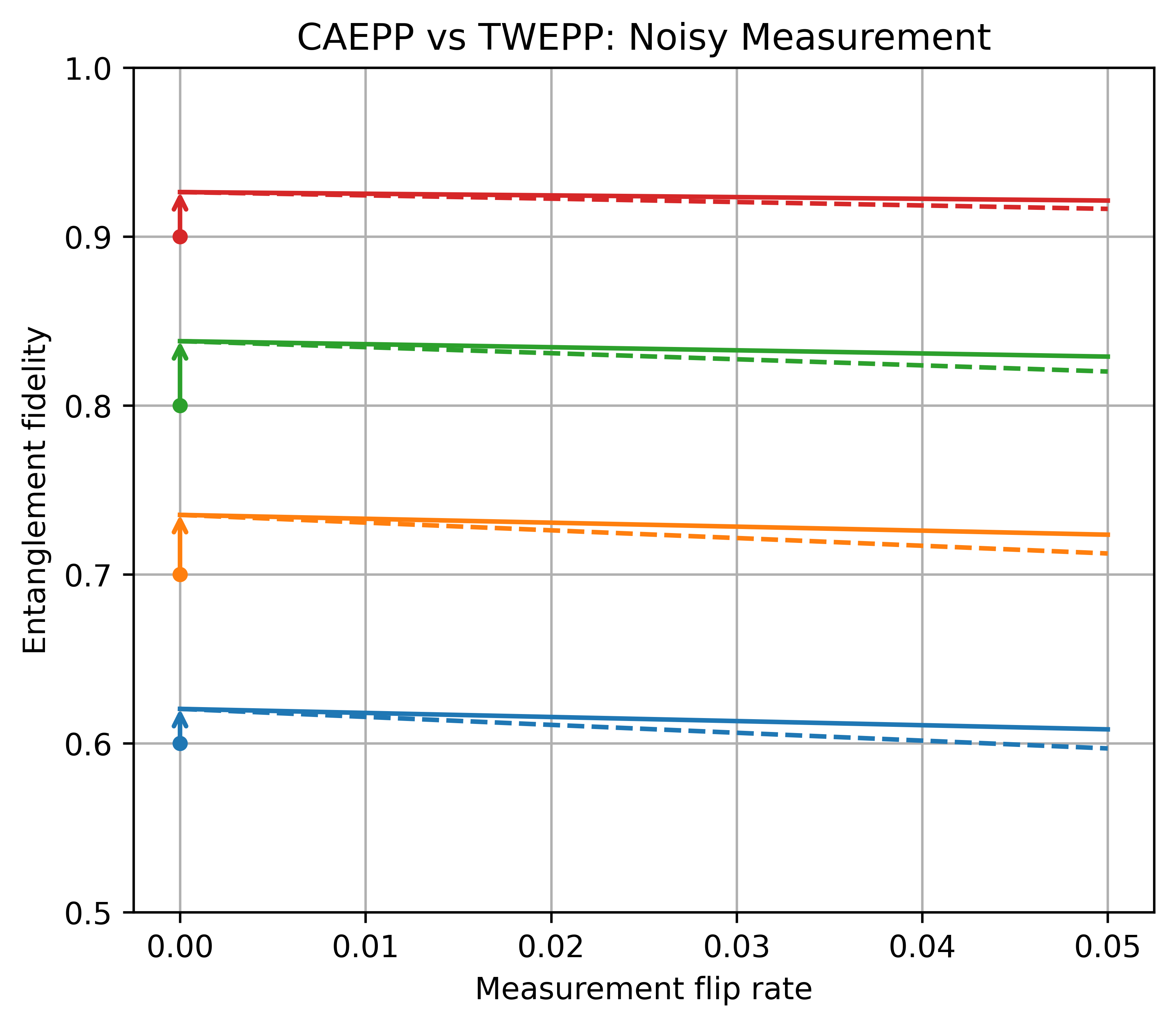}
\subcaption{The entanglement fidelity after a single round \textbf{Success} decreases by a measurement error with a rate $f$.}
\label{fig:meas_error}
\end{minipage}
\hfill
\begin{minipage}{0.32\textwidth}
\centering
\includegraphics[width=\linewidth]{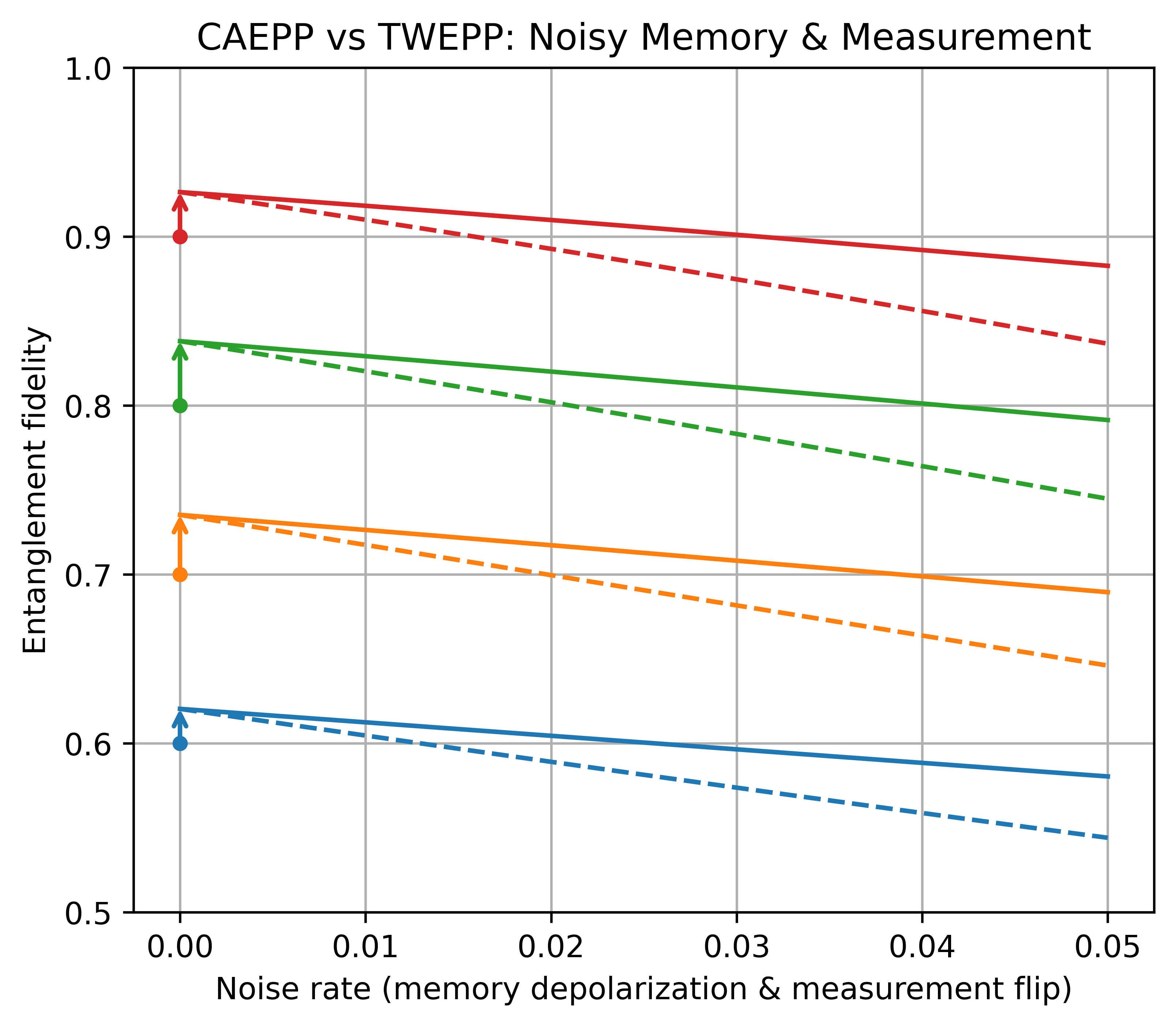}
\subcaption{ The entanglement fidelity after a single round \textbf{Success} decreases by both memory and measurement error with the same rate $e=f$.}
\label{fig:both_error}
\end{minipage}
\caption{ When initial fidelities are $0.6$, $0.7$, $0.8$, and $0.9$, a single round of the CAEPP or TWEPPs in a noiseless scenario equally enhances the fidelities, see the arrows in (a), (b) and (c). When noise is present in (a) quantum memories, (b) single-qubit measurements, and (c) both, the CAEPP (solid) and TWEPPs (dashed) are inequivalent: entanglement enhancements by them are distinct. The comparison above shows that the CAEPP is more robust against noise of quantum memories, {as described in Eq. (\ref{eq:memonoise})}, or single-qubit measurements {in Eq. (\ref{eq:mmnoise})} over TWEPPs.}
\label{fig:noise_comparison}
\end{figure*}

\subsection{Practical advantages of the CAEPP}
The CAEPP and TWEPPs are inequivalent in terms of experimental resources required for their implementations. Table~\ref{tab:caepp-vs-twepp} summarizes distinctions between the CAEPP and TWEPPs in the generalized setting of $m$ carriers and $m+1$ pairs, respectively.

We compare experimental resources required by the CAEPP and TWEPPs, {such as BBPSSW and DEJMPS, and analyze the effect of noise in those resources. Note that for two parties sharing general states in Eq. (\ref{eq:sigmarho}), EP corresponds to an instance of TWEPPs; their implementations require identical experimental resources. The CAEPP, which enhances entanglement as much as EP does, requires a measurement only on Bob's side. Hence, we focus on comparing the experimental resources of TWEPPs and the CAEPP.} 

For a fair comparison, we consider TWEPPs acting on two copies of isotropic states~\eqref{iso}, and the CAEPP on an isotropic state and with a single-qubit carrier through a noisy channel, see also Fig.~\ref{fig:noise_sources}. Noisy resources are listed as follows.

\begin{enumerate}
\item { \textbf{CNOT gate:} Both the CAEPP and TWEPPs share an equal number of CNOT gates. Hence, noise on a CNOT gate affects both the CAEPP and TWEPPs equally. Note that the effect of noisy CNOT has been analyzed ~\cite{Dur1999QuantumRepeaters, briegel1998quantum}. When CNOT gates are noisy, the CAEPP does not show its usefulness over TWEPPs.}

\item \textbf{Quantum memory:}
A qubit stored in a memory may suffer from decoherence. For instance, depolarization noise introduces a noisy channel $\mathcal{D}_e$, by which a state $\rho$ evolves
\begin{align}
\rho \mapsto \mathcal{D}_e(\rho) = (1-e)\rho + e \frac{\I}{2}. \label{eq:memonoise}
\end{align}
We recall that TWEPPs exploit quantum memories for four qubits, whereas the CAEPP for two qubits, see Fig.~\ref{fig:noise_sources}. Then, the effect of noisy quantum memory on the CAEPP and TWEPPs is compared in Fig.~\ref{fig:memory_error}, and the robustness of the CAEPP against noise in quantum memories {in Eq. (\ref{eq:memonoise})} is demonstrated. 
{We recall that the model of noise in Eq. (\ref{eq:memonoise}) has been considered in Refs. \cite{briegel1998quantum, Dur2005, Dur1999QuantumRepeaters}, as a fraction of noise is directly related to entanglement purification. 
} 

\item \textbf{Single-qubit measurement:}
A noiseless measurement is described by POVM elements $\{M_0,M_1\}$ with $M_i=\ket{i}\!\bra{i}$ for $i=0,1$. A noisy measurement has been considered \cite{Dur1999QuantumRepeaters}:
\begin{align}
\widetilde{M}_i &= (1-f)\ket{i}\!\bra{i} + f\,\ket{i+1}\!\bra{i+1}, \label{eq:mmnoise}
\end{align}
where $+$ denotes bitwise addition. TWEPPs apply two single-qubit measurements, whereas the CAEPP does so once, see Fig. \ref{fig:noise_sources}. {We note that the model of noise above has been considered in Refs. \cite{briegel1998quantum, Dur2005, Dur1999QuantumRepeaters, kim2025purification}, as a fraction of noise is directly related to entanglement purification. }

We then recall that the entanglement fidelity increases with noisy measurements \cite{Dur1999QuantumRepeaters}
\bea
F' &= &
\frac{
p_\text{c} \!\left(F^2 + (\frac{1-F}{3})^2\right) + p_\text{i} \!\left(F(\frac{1-F}{3})+(\frac{1-F}{3})^2\right)
}{ \widetilde{P}_{\mathrm{succ}} } \nonumber\\
\widetilde{P}_{\mathrm{succ}} &=&
p_\text{c} \!\left(F^2 + 2F(\frac{1-F}{3}) +
 5(\frac{1-F}{3})^2\right) + \nonumber \\
 &&
4  p_\text{i}  \left(F(\frac{1-F}{3})+(\frac{1-F}{3})^2\right) \nonumber
\eea
where $p_\text{c}$ and $p_\text{i}$ denote the probabilities of correct and erroneous outcomes, respectively. The CAEPP corresponds to the case where  $(p_\text{c},p_\text{i})=(1-f,f)$ since a noisy measurement applies once. TWEPPs have $(p_\text{c},p_\text{i})=((1-f)^2+f^2,\,2f(1-f))$ as they perform noisy measurements at two places. Since $f< 2f(1-f)$ for $f<1/2$, it is clear that the CAEPP is more robust than TWEPPs {against noise in Eq. (\ref{eq:mmnoise})}. For values $f\in[0,0.05]$, the CAEPP and TWEPPs are compared in Fig.~\ref{fig:meas_error} and the robustness of the CAEPP against measurement noise {in Eq. (\ref{eq:mmnoise})} is demonstrated.
\end{enumerate}

\section{Purifying multipartite entanglement} \label{sec:GHZ}  

The CAEPP can be generalized to purification of the GHZ state. Let us describe a tripartite scenario and the CAEPP for GHZ states.

\subsubsection*{Tripartite scenario}
Let us consider a tripartite scenario in which Alice, Bob, and Carol want to share a GHZ state. At \textbf{Initialization}, Alice prepares a GHZ state, sending one qubit to Bob through a Pauli channel $\mathcal{N}_{AB}$ with parameters $(p_{00}, p_{01}, p_{10}, p_{11})$, and another qubit to Carol through a Pauli channel $\mathcal{N}_{AC}$ with parameters $(q_{00}, q_{01}, q_{10}, q_{11})$.  

\subsubsection*{CAEPP for GHZ states}
After initialization, Alice, Bob, and Carol share a GHZ-diagonal state $\sigma$, which is a convex combination of GHZ-basis states:  
\bea
\sigma = \sum_{abc} s_{abc}\, \G_{abc}, \quad \text{where} ~~\G_{abc} = |\G_{abc}\rangle \langle \G_{abc}| \label{ghz1} 
\eea
and
\bea
\ket{\G_{abc}} = Z^c \otimes X^a \otimes X^b \, \frac{1}{\sqrt{2}}(\ket{000}+\ket{111}).\nonumber 
\eea
Note that $s_{abc} = \sum_{k=0,1} p_{ak}\, q_{b,\,k\oplus c}$ for $a,b,c \in \{0,1\}$.

For generality, suppose that Alice, Bob, and Carol share a modified GHZ-diagonal state of the form  
\begin{align}
\rho &= \sum_{abc} r_{abc}\, \G_{abc}. \label{ghz2}
\end{align}

The CAEPP in a tripartite system proceeds as follows.  
Alice prepares two carrier qubits in the state $\ket{0}$, applies a CNOT gate to each, and sends one carrier to Bob and the other to Carol.  
Bob and Carol each apply a CNOT gate followed by a measurement on their respective carrier qubits.
After classical communication, the state is kept only if both outcomes are $0$.  
When the protocol succeeds, the unnormalized output state is  
\begin{align}
\sum_{abck} r_{abk}\, s_{a,b,k\oplus c}\, \G_{abc}.
\end{align}
The success probability and the fidelity with $\G_{000}$ are then given by  
\begin{align}
p_{\mathrm{succ}} &= \sum_{abck} r_{abk}\, s_{a,b,k\oplus c}, \\
F' &= \frac{r_{000} s_{000} + r_{001} s_{001}}{p_{\mathrm{succ}}}.
\end{align}
{ The resulting fidelity above can be compared with the condition in Eq. (\ref{eq:condition}). Thus, we have shown that the CAEPP works for distilling tripartite GHZ states. The construction may also be applied to $n$-partite GHZ states for $n>3$.}



\section{Conclusion and discussion} \label{sec:conclusion}  

\subsubsection*{Summary}  
We have introduced carrier-assisted entanglement purification (CAEPP), a protocol that purifies shared entanglement without consuming additional pairs of entangled states. The central idea is to replace auxiliary entangled pairs with one or more transmitted carrier qubits, thereby reducing memory requirements and noisy measurements while retaining purification capability. { The protocol shares the practical usefulness with TWEPPs by tolerating higher noise and also contains practical advantages over TWEPPs, requiring minimal experimental resources, such as two quantum memories and single-qubit transmission. }

We presented the CAEPP with a single-qubit carrier. The protocol enables two parties to achieve an ebit in two rounds of the protocol if a carrier is sent through a noiseless channel. When a channel for a carrier qubit is noisy, the CAEPP contains limitations in the purification; although the fidelity may increase, it converges to a maximum value, called maximum convergent fidelity. We have investigated the CAEPP against noisy qubit channels of a single-qubit carrier. 

We then showed how the mCAEPP with stabilizer codes can resolve the limitation; the protocol has the maximum convergent fidelity as $1$. Examples with two, three, and four carriers using different stabilizer codes are presented. We further proved that the maximum convergent fidelity approaches $1$ with multiple carriers, showing that CAEPP can generally enable parties to share ebits through noisy channels. 

The CAEPP contains practical advantages over TWEPPs. The CAEPP requires only one shared pair and one transmitted carrier, in contrast to TWEPPs that consume multiple pre-shared pairs and require a substantial use of quantum memory. In addition, the CAEPP applies a smaller number of measurements than TWEPPs. As a result, the CAEPP is more resilient to memory noise and measurement errors. 

Finally, we have demonstrated the purification of multipartite entanglement exemplified with a tripartite GHZ state. {We leave it to future works to establish thresholds, yields, or convergence for general multipartite stabilizer or non-stabilizer states. Distillation of GHZ states shares many features in common with stabilizer distillation that works for the purification of stabilizer states \cite{Matsumoto2003Conversion, GlancyKnillVasconcelos2006}. Note also that in multipartite quantum systems, there are entangled states that are not equivalent under stochastic LOCC; hence, there is no unique unit of multipartite entanglement. For instance, GHZ and W states are not interconvertible. It would be interesting to find the CAEPP for purifying multipartite entangled states in general.}

\subsubsection*{Discussion} 
In future directions, it would be interesting to investigate the CAEPP for creating entanglement beyond the purification of entanglement. Note that the CAEPP shares some similarities with the entanglement distribution with separable states (EDSS) protocol~\cite{fedrizzi2013distributing}, where entanglement is established through carrier transmission from initially separable states. 
 
It would also be interesting to develop the mCAEPP by optimizing \emph{adaptive stabilizer strategies}, which depend on a choice of stabilizer codes. Identifying adaptive schemes for an efficient mCAEPP is a key to realizing the purification of entanglement in practical realizations over noisy channels. Clearly, it is also of fundamental interest to consider the CAEPP for high-dimensional quantum systems. 

We envisage that the CAEPP provides a practical and robust alternative to TWEPPs. 
{The CAEPP reduces memory demands and can tolerate higher rates of measurement noise, and thus offers the pathway toward scalable entanglement distribution in near-term quantum networks. We leave it as future investigations to take real hardware budgets into account, such as wall-clock time, channel loss, dark counts, reset time, memory lifetime, or feed-forward latency, to realize entanglement purification in real environments.}

\section*{Acknowledgement}

This work is supported by the National Research Foundation (RS-2024-00408613) and the Institute for Information \& Communication Technology Promotion (IITP) (RS-2023-00229524, RS-2025-02304540, RS-2025-25464876, RS-2025-25464616). 
J.K.\ was supported in part by the Danish National Research Foundation (DNRF) through the Center of Excellence CLASSIQUE, Grant No.\ DNRF187.

\appendix

{

\section{Coefficient update and the fixed point}  \label{App}

This appendix establishes the result stated in Section~\ref{sec:fidelity_to_one}.
We proceed in three steps:
(i) we express one purification round as a normalized linear map on the Bell coefficients;
(ii) we prove convergence to a unique fixed point; and
(iii) we show that the fixed-point fidelity approaches unity exponentially in the number of carriers for a depolarizing channel with $p_{00} \in (\tfrac{1}{2}, 1)$.
Note that the case $p_{00} = 1$ is trivial, since the initial pair is $\ket{\phi^+}$ and does not require purification.
{The proof applies generally to any Pauli channel with $p_{00}>\tfrac12$ by incorporating the random-unitary carrier coding described in Section~\ref{sec:channel_twirling}.}

\subsection{One-round update as a normalized linear map}
We derive the fixed-point equations satisfied by the shared state at convergence and show that the fixed-point fidelity $F_\star$ (i.e., the fidelity bound) approaches $1$ exponentially as the number of carriers $m$ increases, under a depolarizing channel~\eqref{depol_chan} with $p := p_{00} \in (\tfrac{1}{2},1)$.
We employ the check operators
\begin{align}
\{X_1 X_m, \ldots, X_{m-1} X_m, Z_0 Z_1 \cdots Z_m\}, \label{m-stab}
\end{align}
where $m \ge 2$. For $m=1$, it is reduced to the check operators $\{Z_0 Z_1\}$ implemented by CNOT gates.

Let the shared state after Round $n$ be a Bell-diagonal state~\eqref{qBDS} with coefficients $\bm{q}^{(n)} = (q_{00}, q_{01}, q_{10}, q_{11})$.  
At the \textbf{Pre-processing} step, Alice and Bob apply $R_X(+\tfrac{\pi}{2}) \otimes R_X(-\tfrac{\pi}{2})$, which rearranges the coefficients to
\begin{align}
\bm{r}^{(n)} = (r_{00}, r_{01}, r_{10}, r_{11}) = (q_{00}, q_{11}, q_{10}, q_{01}).
\label{rotatedq}
\end{align}

The errors affecting the shared state and the $m$ carriers during transmission can be represented by a $2m{+}2$-bit \emph{error string}  
\begin{align}
X^{x_0} Z^{z_0} \otimes X^{x_1} Z^{z_1} \otimes \cdots \otimes X^{x_m} Z^{z_m}, \label{error_string}
\end{align}
up to a global phase.  
Here, $(x_0, z_0)$ corresponds to the shared pair, while $(x_k, z_k)$ for $k \in \{1,\ldots,m\}$ corresponds to the $k$-th carrier qubit.  

The error probabilities of the input state are given by $\bm{r}^{(n)}$~\eqref{rotatedq}, while the error probabilities of carrier qubits are distributed as
\begin{align}
P(x_k, z_k) = 
\begin{cases}
p, & (x_k, z_k) = (0,0), \\
\tfrac{1-p}{3}, & (x_k, z_k) \in \{(0,1),(1,0),(1,1)\},
\end{cases} \label{error_prob}
\end{align}
with
\begin{align}
\alpha := \E[(-1)^{x_k}] = \E[(-1)^{z_k}] = \E[(-1)^{x_k+z_k}] = \frac{4p-1}{3}.
\end{align}

A purification round succeeds exactly when the error string commutes with all check operators of~\eqref{m-stab}, which is equivalent to the syndrome constraints
\begin{align}
\bigoplus_{k=0}^{m} x_k = 0, 
\qquad 
z_1 = \cdots = z_m. \label{survival}
\end{align}
For notational convenience, we introduce the Boolean predicate delta
\[
\delta[\mathsf{P}] :=
\begin{cases}
1, & \text{if the predicate $\mathsf{P}$ is true},\\
0, & \text{otherwise}.
\end{cases}
\]

Define the elementary success constraints
\begin{align}
\delta_X &:= \delta\!\left[\bigoplus_{k=0}^m x_k = 0\right]
      = \frac12\left(1+\prod_{k=0}^m(-1)^{x_k}\right), \label{deltaX}\\
\delta_Z &:= \delta[z_1=\cdots=z_m]
      = \prod_{k=1}^{m}\frac{1+(-1)^{z_k}}{2}
        +\prod_{k=1}^{m}\frac{1-(-1)^{z_k}}{2}. \label{deltaZ}
\end{align}
Then the success indicator is simply
\begin{align}
\delta_{\mathrm{succ}} := \delta_X\,\delta_Z. \label{succ_ind}
\end{align}

Since errors on different carriers are independent and expectations of products of independent variables factorize, we set
\begin{align}
A_m &:= \E\!\left[\prod_{k=1}^m \tfrac{1+(-1)^{z_k}}{2} \right] = \left(\tfrac{1+\alpha}{2}\right)^m, \\
B_m &:= \E\!\left[\prod_{k=1}^m \tfrac{1-(-1)^{z_k}}{2} \right] = \left(\tfrac{1-\alpha}{2}\right)^m, \\
C_m &:= \E\!\left[\prod_{k=1}^m \tfrac{(-1)^{x_k}+(-1)^{x_k+z_k}}{2} \right] = \alpha^m,
\end{align}
with $\E[\tfrac{(-1)^{x_k}-(-1)^{x_k+z_k}}{2}] = 0$.

We now express everything back in $\bm{q}^{(n)}$ using \eqref{rotatedq}. The success probability then evaluates to
\begin{align}
P_{\mathrm{succ}} = \E[\delta_{\mathrm{succ}}]
= \tfrac{1}{2} \left(A_m + B_m + (q_{00} + q_{11} - q_{10} - q_{01}) C_m\right). \label{Psucc}
\end{align}

Similarly, to select the output Bell label $(s,t)$, define
\begin{align}
\delta_s &:= \delta[s=x_0] = \frac{1+(-1)^{s+x_0}}{2}, \label{deltas}
\end{align}
and, under the success constraint $\delta_Z=1$ (hence $z_1=\cdots=z_m$), the phase label obeys
$t = z_0 \oplus z_1$. Accordingly, define
\begin{align}
\delta_t &:= \delta[t=z_0\oplus z_1] 
= \frac{1+(-1)^{t+z_0+z_1}}{2}. \label{deltat}
\end{align}
Then the Bell indicators factor as
\begin{align}
\delta_{st} := \delta_{\mathrm{succ}}\,\delta_s\,\delta_t, \qquad
q'_{st}=\frac{\E[\delta_{st}]}{P_{\mathrm{succ}}}. \label{Bell_ind}
\end{align}

The updated Bell coefficients then evaluate to
\begin{align}
q'_{00} & = \frac{(A_m+C_m) q_{00} + B_m q_{11}}{2P_{\mathrm{succ}}}, \label{q00} \\
q'_{01} & = \frac{(A_m+C_m) q_{11} + B_m q_{00}}{2P_{\mathrm{succ}}}, \label{q01} \\
q'_{10} & = \frac{(A_m-C_m) q_{10} + B_m q_{01}}{2P_{\mathrm{succ}}}, \label{q10} \\
q'_{11} & = \frac{(A_m-C_m) q_{01} + B_m q_{10}}{2P_{\mathrm{succ}}}, \label{q11}
\end{align}
where
Equations~\eqref{q00}–\eqref{q11} can be written compactly as a normalized linear update
\begin{align}
\bm q^{(n+1)} &= \frac{L_m \bm q^{(n)}}{\mathbf 1^\top L_m \bm q^{(n)}}, \label{linear_update} \\
L_m &:= \frac{1}{2}
\begin{pmatrix}
A_m{+}C_m & 0 & 0 & B_m\\
B_m & 0 & 0 & A_m{+}C_m\\
0 & B_m & A_m{-}C_m & 0\\
0 & A_m{-}C_m & B_m & 0
\end{pmatrix},
\end{align}
where $\mathbf 1=(1,1,1,1)^\top$ and $\mathbf 1^\top L_m \bm q^{(n)} = P_{\mathrm{succ}}$.

\subsection{Convergence to a unique fixed point}
We now prove that iteration~\eqref{linear_update} converges to a unique fixed point.

For $p\in(\tfrac12,1)$, we have $\alpha\in(0,1)$ and therefore $A_m>C_m>0$ and $B_m>0$.
A direct computation shows that every entry of $L_m^2$ is strictly positive. Hence, $L_m$ is a \emph{primitive nonnegative matrix}.

By the Perron--Frobenius theorem for primitive nonnegative matrices~\cite{Seneta2006}, $L_m$ has a unique dominant eigenvalue $\lambda_{\max}>0$ and corresponding right and left eigenvectors $\bm v > 0$ and $\bm w > 0$ such that $L_m \bm v=\lambda_{\max}\bm v$, $\bm w^\top L_m=\lambda_{\max}\bm w^\top$, and $\bm w^\top \bm v = 1$.
Moreover,
\begin{align}
\left(\frac{L_m}{\lambda_{\max}}\right)^n \xrightarrow[]{n\to\infty} \bm v \bm w^\top,
\end{align}
and the convergence is exponentially fast, i.e., subdominant modes decay as $|\lambda_i/\lambda_{\max}|^n$.
Since $\bm w^\top \bm q^{(0)}>0$, it follows that
\begin{align}
\frac{L_m^n \bm q^{(0)}}{\mathbf 1^\top L_m^n \bm q^{(0)}} \xrightarrow[n\to\infty]{}
\frac{\bm v}{\mathbf 1^\top \bm v}. \label{PF_normalized_limit}
\end{align}
By induction on $n$, iteration of the normalized update~\eqref{linear_update} satisfies $\bm q^{(n)} = L_m^n \bm q^{(0)} / \mathbf{1}^\top L_m^n \bm q^{(0)}$. Eq.~\eqref{PF_normalized_limit} directly implies that $\bm q^{(n)}$ converges to the unique fixed point $\bm q^\star = \bm v/(\mathbf 1^\top \bm v)$, independently of the initial Bell-diagonal coefficients $\bm q^{(0)}$.

\subsection{Fixed point fidelity converges to $1$}
At convergence, the coefficients are fixed points: $\bm{q}^\star = L_m \bm q^\star / \mathbf{1}^\top L_m \bm q^\star$.

From \eqref{q00} and \eqref{Psucc},  
\begin{align}
(B_m - 2C_m(q^\star_{10}+q^\star_{01})) q^\star_{00} = B_m q^\star_{11}, \lb
q^\star_{10}+q^\star_{01} = \tfrac{B_m}{2C_m} \Bigl(1- \tfrac{q^\star_{11}}{q^\star_{00}}\Bigr) 
   \le \tfrac{B_m}{2C_m}. \label{qbound1}
\end{align}
Next, using \eqref{q01}, \eqref{Psucc}, and the bound in \eqref{qbound1}, we obtain
\begin{align}
2 P_{\mathrm{succ}} q^\star_{01} &=  (A_m+C_m) q^\star_{11} + B_m q^\star_{00} 
   \;\;\ge\;\; (A_m+C_m) q^\star_{11}, \lb
q^\star_{11} &\le \tfrac{2 P_{\mathrm{succ}}}{A_m+C_m} q^\star_{01}. 
\end{align}
Since $2 P_{\mathrm{succ}} \le A_m+B_m+C_m$, we can further bound
\begin{align}
q^\star_{11} \le \tfrac{A_m+B_m+C_m}{A_m+C_m} q^\star_{01}. 
\end{align}
Noticing $\tfrac{A_m+B_m+C_m}{A_m+C_m} \le 2$ and $q^\star_{01} \le q^\star_{10}+q^\star_{01}$ and applying \eqref{qbound1}, we conclude
\begin{align}
q^\star_{11} \le 2(q^\star_{10}+q^\star_{01}) \le \tfrac{B_m}{C_m}. \label{qbound2}
\end{align}

Finally, when $\tfrac{1}{2}<p<1$,
\begin{align}
\frac{B_m}{C_m} &= \Bigl(\frac{1-\alpha}{2\alpha}\Bigr)^m = \Bigl(\frac{2(1-p)}{4p-1}\Bigr)^m \xrightarrow{m \to \infty} 0, \label{bc}
\end{align}
Combining \eqref{qbound1}, \eqref{qbound2}, and \eqref{bc} gives
\begin{align}
1-q^\star_{00} = q^\star_{01} + q^\star_{10} + q^\star_{11} \le \tfrac{3B_m}{2C_m} \xrightarrow{m \to \infty} 0.
\end{align}
Thus, the fixed-point fidelity $F_\star = q^\star_{00}$ converges exponentially to $1$ as $m$ increases, completing the proof.
}

\bibliographystyle{IEEEtran}
\bibliography{reference}

\end{document}